\documentclass[12pt]{article}

\usepackage{geometry}
\usepackage{graphicx}
\usepackage[utf8]{inputenc}
\usepackage{amsmath,amscd}    
\usepackage{graphicx}   
\usepackage{verbatim}   
\usepackage{color}      
\usepackage{subfigure}  
\usepackage{hyperref}   
\usepackage{bbm}
\usepackage{amsfonts}
\usepackage{mathtools}
\usepackage{amssymb}
\usepackage{tensor}

\begin{document}
\author{
   Adam B. Gillard\\
  \and 
  Niels G. Gresnigt\footnote{Corresponding author: niels.gresnigt@xjtlu.edu.cn \newline \indent \;\;\;Department of Mathematical Sciences, Xi'an Jiaotong-Liverpool University, 111 Ren'ai Road, \indent\;\;\;Suzhou HET, Jiangsu, China 215123.}
}
\title{The $C\ell(8)$ algebra of three fermion generations with spin and full internal symmetries}
\maketitle

\begin{abstract}
In this paper, the basis states of the minimal left ideals of the complex Clifford algebra $C\ell(8)$ are shown to contain three generations of Standard Model fermion states, with full Lorentzian, right and left chiral, weak isospin, spin, and electrocolor degrees of freedom. The left adjoint action algebra of $C\ell(8)\cong\mathbb{C}(16)$ on its minimal left ideals contains the Dirac algebra, weak isopin and spin transformations. The right adjoint action algebra on the other hand encodes the electrocolor symmetries. These results extend earlier work in the literature that shows that the eight minimal left ideals of $\mathbb{C}(8)\cong C\ell(6)$ contain the quark and lepton states of one generation of fixed spin. Including spin degrees of freedom extends $C\ell(6)$ to $C\ell(8)$, which unlike $C\ell(6)$ admits a triality automorphism. It is this triality that underlies the extension from a single generation of fermions to exactly three generations. 
\end{abstract}



%
%
%
%
%
\section{Introduction}
The spinors of the complex Clifford algebra $C\ell(8)$ are used to represent three full generations of Standard Model (SM) fermion states including spin and full SM internal symmetries. It was shown in \cite{Stoica2018} that the eight minimal left ideals of the algebra of $8\times8$ complex matrices $\mathbb{C}(8)\cong C\ell(6)$ contains the 64 elementary fermion states of one generation of fixed spin, inclusive of antiparticles. In this paper we show that including spin degrees of freedom extends $C\ell(6)$ to $C\ell(8)$. This is the first main result of the present paper. We then show that via the triality automorphism, which is present in $C\ell(8)$ but not in $C\ell(6)$, we obtain exactly three full generations of fermion states, rather than a single generation as with $C\ell(6)$. This is the second main result of this paper.

The sixteen minimal left ideals of the $2^8$ $\mathbb{C}$-dimensional Clifford algebra $C\ell(8)$ can be represented by sixteen column vectors in the isomorphic matrix algebra $\mathbb{C}(16)$ of $16\times16$ matrices over the complex numbers $\mathbb{C}$. The action of left and right matrix multiplication differs. Left multiplication of a minimal left ideal column matrix\footnote{i.e. a square matrix with one nonzero column} interchanges rows, and hence produces transformations within the minimal left ideals themselves. In contrast, right multiplication of a minimal left ideal by an arbitrary matrix in $\mathbb{C}(16)$ interchanges columns, and hence transforms between different minimal left ideals. Symbolically we write:
\begin{equation}
C\ell(8)\triangleright\{P_1,\cdots,P_{16}\}\triangleleft C\ell(8),\quad\leftrightarrow \quad\mathbb{C}(16)\triangleright\{\rho(P_1),\cdots,\rho(P_{16})\}\triangleleft\mathbb{C}(16),
\end{equation} 
where $P_i\in C\ell(8)$ are the algebraic minimal left ideals and the $\rho(P_i)\in\mathbb{C}(16)$ are the corresponding column vector matrix representations of $P_i$ that come about through the isomorphism $C\ell(8)\cong\mathbb{C}(16)$.

In this paper we interpret the basis states of these minimal left ideals as SM leptons and quarks. Doing so requires that the left and right adjoint actions of $C\ell(8)$ be interpreted physically. Such an interpretation for one generation of fermions with fixed spin in terms of $C\ell(6)$, and its matrix isomorphism with $\mathbb{C}(8)$, was carried out in \cite{Stoica2018}. The left adjoint action algebra $C\ell(6)_L$ in that case turns out physically to correspond to the tensor product $C\ell(4)_{\mathrm{Dirac}}\otimes_{\mathbb{C}}C\ell(2)_{\mathrm{isospin}}$, allowing for the identification of basis states of an individual minimal left ideal with Lorentzian chiral indices. The action of the Dirac algebra $C\ell(4)_{\mathrm{Dirac}}$ is reducible on $\mathbb{C}(8)$, and permutes the rows of each ideal. It furthermore naturally splits each ideal into two four $\mathbb{C}$-dimensional spinors, whose left chiral components are permuted by the $SU(2)_L$ symmetry. The right adjoint action algebra $C\ell(6)_R$ on the other permutes between different ideals. Its maximal totally isotropic subspace (MTIS) symmetry group is $U(3)$, under which the ideals transform as a color singlet, triplet, antisinglet, and antitriplet. The right adjoint actions therefore account for the $U(3)_{\mathrm{ec}}$ unbroken electrocolor gauge symmetries. A single generation of leptons and quarks with the desired SM symmetries can therefore be represented in terms of the basis states of the minimal left ideals of $C\ell(6)$.  

We show that the extension of the above via the inclusion of spin degrees of freedom, represented by the algebra $C\ell(2)_{\mathrm{spin}}$ generated from the $su(2)_{\textrm{spin}}$ generators, enlarges the left adjoint action algebra from $C\ell(6)_L$ to $C\ell(8)_L\cong \mathbb{C}(16)$, so that:
\begin{equation}
C\ell(8)\triangleright\{P_1,\cdots, P_{16}\},\;\leftrightarrow\; \mathbb{C}(16)\triangleright\{\rho(P_1),\cdots,\rho(P_{16})\}.
\end{equation}
The first eight rows of each minimal left ideal column corresponds to spin-up states while the second set of eight rows in each minimal left ideal column corresponds to spin-down states. The minimal left ideals of $C\ell(8)$ are algebraic pinors. The algebraic spinors, which correspond to the physical states, are obtained by disregarding the odd Clifford grade basis states. These spinors therefore reside in $C\ell^+(8)\cong C\ell(7)\cong\mathbb{C}(8)\oplus\mathbb{C}(8)$\footnote{That is, a $C\ell(8)$ spinor is isomoprhic to a $C\ell(7)$ pinor.}, and within each $8\times8$ block one finds what looks like one generation of elementary fermion states of fixed spin. The left action subalgebra $C\ell(7)_{\mathrm{Dirac+isospin+spin-z}}$, which is $C\ell(6)_{\textrm{Dirac+isospin}}$ taken together with the spin-$z$ operator acts irreducibly on $8\times8$ blocks, so that its action on $\mathbb{C}(8)$ extends to an action on $\mathbb{C}(8)\oplus\mathbb{C}(8)$ via the direct sum. However the ladder spin operators in $C\ell(2)_{\mathrm{spin}}$ do not preserve spinors, mapping them to pinors, and therefore to unphysical states. 

The right action algebra on the other hand is the MTIS symmetry group of $C\ell(8)$. This is the group $U(4)$. The electrocolor subgroup $U(3)_{\mathrm{ec}}\subset U(4)$ acts in a direct sum manner, reducibly on $8\times8$ blocks, thus preserving the color charges of each spinor. The remaining $U(4)$ transformations map spinors into pinors. Therefore, the restriction of MTIS symmetries that act on physical spinors is the electrocolor group $U(3)_{\mathrm{ec}}$, acting irreducibly on each $\mathbb{C}(8)$ block in $\mathbb{C}(8)\oplus\mathbb{C}(8)$. 

A natural question is how to best extend these results from a single generation to three. Although it is possible to represent three generations of fermion states of fixed spin within a single copy of $C\ell(6)$ \cite{furey2014generations,furey2018three}, in that case the physical states are no longer the basis states of minimal left ideals. Others have considered the exceptional Jordan algebra   $J_3(\mathbb{O})$ as a natural mathematical structure to describe three generations of fermions \cite{dubois2016exceptional,todorov2018deducing,dubois2019exceptional}. Our approach presented here is different. Unlike $C\ell(6)$, the spin-extended algebra $C\ell(8)$ admits a triality automorphism of its associated group $\textrm{Spin}(8)$. The two spinor representations and the fundamental vector representation of this group are all eight-dimensional. Triality, which is a non-linear automorphism of order three, permutes between these representations. We show that it is this automorphism that extends our results from a single generation of fermions to exactly three.

The study of $C\ell(8)$ spinors as physical quantum states is natural. The Hilbert space forms a tensor algebra of states quotiented such that there is a norm. It then follows that one must have a normed division algebra, of which there are only four \cite{hurwitz1922komposition,zorn1931theorie}. Normed division algebras imply normed trialities, with associated spinor spaces restricted to $\mathbb{C}$-dimensions $n=1,2,4$ and 8 \cite{baez2002octonions}. The role of division algebras and Clifford algebras in particle physics has some history. In the 1970s the octonions were related to the color symmetries of a generation of quarks \cite{gunaydin1973quark,gunaydin1974quark}. More recently there has been a revived interest in using division algebras to construct a theoretical basis for the SM gauge groups and the observed particle spectrum \cite{trayling2001geometric,schmeikal2001minimal,schmeikal2005algebra,trayling2004cl,dixon2013division,furey2016standard,gresnigt2018braids,gresnigt2019braided,dray2010octonionic,catto2018quantum, burdik2018hurwitz,perelman2019r}. For example, in \cite{furey2016standard}, the algebra $C\ell(6)$, that also forms the basis of \cite{Stoica2018}, is derived from the left adjoint actions of the complex octonions $\mathbb{C}\otimes\mathbb{O}$ on themselves. 

Sec.\ref{sec:onegen} reviews the earlier work \cite{Stoica2018}, of which the present paper constitutes a natural extension. In Sec.\ref{sec:DWIS} we show how the inclusion of spin extends  $C\ell(6)$ to $C\ell(8)$. In Sec.\ref{sec:MTIS} the $C\ell(8)$ pinors and the right acting MTIS symmetry group $U(4)$ are set up, before demonstrating the emergence of three generations from the triality automorphism in Sec.\ref{sec:3gens}. Finally, in Sec.\ref{sec:sed} the relationship between the results of the present paper with those of an earlier three generation model in terms of complex sedenions is discussed. 

\section{One generation of fermions from $\mathbb{C}(8)\cong C\ell(6)$}
\label{sec:onegen}
We start with a brief review of some of the main results in \cite{Stoica2018}, upon which the present paper builds. The full details can be found in that work.

Starting with a Witt basis $\{q_i,q_i^\dagger\}$, $i=1,2,3$ of $C\ell(6)$ generators that satisfy the conditions:
\begin{equation}
q_i^2=(q_i^\dagger)^2=\{q_i,q_j\}=\{q_i^\dagger,q_j^\dagger\}=0,\quad \{q_i,q_j^\dagger\}=\delta_{ij},
\end{equation}
together with a ``vacuum'' idempotent state $\mathfrak{p}\equiv q_1q_2q_3q_3^\dagger q_2^\dagger q_1^\dagger$ satisfying $\mathfrak{p}^2=\mathfrak{p}$ and $q_i\mathfrak{p}=0$, a minimal left ideal $P_1\in C\ell(6)$ can be constructed in terms of the eight-complex-dimensional basis\footnote{As a shorthand notation we have $q_{ij}^\dagger\equiv q_i^\dagger q_j^\dagger$ and $q_{ijk}\equiv q_iq_jq_k$ e.t.c.}:
\begin{equation}
P_1:\quad \{\mathfrak{p},q_{23}^\dagger \mathfrak{p},q_{31}^\dagger\mathfrak{p},q_{12}^\dagger\mathfrak{p},q_{123}^\dagger\mathfrak{p},q_1^\dagger\mathfrak{p},q_2^\dagger\mathfrak{p},q_3^\dagger\mathfrak{p}\}.
\end{equation}
The remaining seven minimal left ideals $P_2,\cdots,P_8$ are likewise constructed with the following bases:
\begin{eqnarray}
\nonumber P_2:&\quad& \{\mathfrak{p},q_{23}^\dagger \mathfrak{p},q_{31}^\dagger\mathfrak{p},q_{12}^\dagger\mathfrak{p},q_{123}^\dagger\mathfrak{p},q_1^\dagger\mathfrak{p},q_2^\dagger\mathfrak{p},q_3^\dagger\mathfrak{p}\}q_{32}, \\
\nonumber P_3:&\quad& \{\mathfrak{p},q_{23}^\dagger \mathfrak{p},q_{31}^\dagger\mathfrak{p},q_{12}^\dagger\mathfrak{p},q_{123}^\dagger\mathfrak{p},q_1^\dagger\mathfrak{p},q_2^\dagger\mathfrak{p},q_3^\dagger\mathfrak{p}\}q_{13}, \\
\nonumber P_4:&\quad& \{\mathfrak{p},q_{23}^\dagger \mathfrak{p},q_{31}^\dagger\mathfrak{p},q_{12}^\dagger\mathfrak{p},q_{123}^\dagger\mathfrak{p},q_1^\dagger\mathfrak{p},q_2^\dagger\mathfrak{p},q_3^\dagger\mathfrak{p}\}q_{21}, \\
P_5:&\quad& \{\mathfrak{p},q_{23}^\dagger \mathfrak{p},q_{31}^\dagger\mathfrak{p},q_{12}^\dagger\mathfrak{p},q_{123}^\dagger\mathfrak{p},q_1^\dagger\mathfrak{p},q_2^\dagger\mathfrak{p},q_3^\dagger\mathfrak{p}\}q_{321}, \\
\nonumber P_6:&\quad& \{\mathfrak{p},q_{23}^\dagger \mathfrak{p},q_{31}^\dagger\mathfrak{p},q_{12}^\dagger\mathfrak{p},q_{123}^\dagger\mathfrak{p},q_1^\dagger\mathfrak{p},q_2^\dagger\mathfrak{p},q_3^\dagger\mathfrak{p}\}q_{1}, \\
\nonumber P_7:&\quad& \{\mathfrak{p},q_{23}^\dagger \mathfrak{p},q_{31}^\dagger\mathfrak{p},q_{12}^\dagger\mathfrak{p},q_{123}^\dagger\mathfrak{p},q_1^\dagger\mathfrak{p},q_2^\dagger\mathfrak{p},q_3^\dagger\mathfrak{p}\}q_{2}, \\
\nonumber P_8:&\quad& \{\mathfrak{p},q_{23}^\dagger \mathfrak{p},q_{31}^\dagger\mathfrak{p},q_{12}^\dagger\mathfrak{p},q_{123}^\dagger\mathfrak{p},q_1^\dagger\mathfrak{p},q_2^\dagger\mathfrak{p},q_3^\dagger\mathfrak{p}\}q_{3}.
\end{eqnarray}
One can subsequently write down a $C\ell(6)\cong\mathbb{C}(8)$ space of states in terms of the complex-valued $8\times8$ matrix:
\begin{equation}
(\rho(P_1),\rho(P_2),\rho(P_3),\rho(P_4),\rho(P_5),\rho(P_6),\rho(P_7),\rho(P_8))
\end{equation} 
where $P_i\in C\ell(6)\cong\mathbb{C}(8)\ni\rho(P_i)$. Symbolically on the left, and literally on the right we can express $\sum_{i=1}^8\rho(P_i)$ as
\begin{equation}
\nonumber\left(\begin{smallmatrix}
1\mathfrak{p}1 & 1\mathfrak{p}q_{23} & 1\mathfrak{p}q_{31} & 1\mathfrak{p}q_{12} & 1\mathfrak{p}q_{321} & 1\mathfrak{p}q_1 & 1\mathfrak{p}q_2 & 1\mathfrak{p}q_3 \\
q_{23}^\dagger\mathfrak{p}1 & q_{23}^\dagger\mathfrak{p}q_{23} & q_{23}^\dagger\mathfrak{p}q_{31} & q_{23}^\dagger\mathfrak{p}q_{12} & q_{23}^\dagger\mathfrak{p}q_{321} & q_{23}^\dagger\mathfrak{p}q_1 & q_{23}^\dagger\mathfrak{p}q_2 & q_{23}^\dagger\mathfrak{p}q_3 \\
q_{31}^\dagger\mathfrak{p}1 & q_{31}^\dagger\mathfrak{p}q_{23} & q_{31}^\dagger\mathfrak{p}q_{31} & q_{31}^\dagger\mathfrak{p}q_{12} & q_{31}^\dagger\mathfrak{p}q_{321} & q_{31}^\dagger\mathfrak{p}q_1 & q_{31}^\dagger\mathfrak{p}q_2 & q_{31}^\dagger\mathfrak{p}q_3 \\
q_{12}^\dagger\mathfrak{p}1 & q_{12}^\dagger\mathfrak{p}q_{23} & q_{12}^\dagger\mathfrak{p}q_{31} & q_{12}^\dagger\mathfrak{p}q_{12} & q_{12}^\dagger\mathfrak{p}q_{321} & q_{12}^\dagger\mathfrak{p}q_1 & q_{12}^\dagger\mathfrak{p}q_2 & q_{12}^\dagger\mathfrak{p}q_3 \\
q_{321}^\dagger\mathfrak{p}1 & q_{321}^\dagger\mathfrak{p}q_{23} & q_{321}^\dagger\mathfrak{p}q_{31} & q_{321}^\dagger\mathfrak{p}q_{12} & q_{321}^\dagger\mathfrak{p}q_{321} & q_{321}^\dagger\mathfrak{p}q_1 & q_{321}^\dagger\mathfrak{p}q_2 & q_{321}^\dagger\mathfrak{p}q_3 \\
q_{1}^\dagger\mathfrak{p}1 & q_{1}^\dagger\mathfrak{p}q_{23} & q_{1}^\dagger\mathfrak{p}q_{31} & q_{1}^\dagger\mathfrak{p}q_{12} & q_{1}^\dagger\mathfrak{p}q_{321} & q_{1}^\dagger\mathfrak{p}q_1 & q_{1}^\dagger\mathfrak{p}q_2 & q_{1}^\dagger\mathfrak{p}q_3 \\
q_{2}^\dagger\mathfrak{p}1 & q_{2}^\dagger\mathfrak{p}q_{23} & q_{2}^\dagger\mathfrak{p}q_{31} & q_{2}^\dagger\mathfrak{p}q_{12} & q_{2}^\dagger\mathfrak{p}q_{321} & q_{2}^\dagger\mathfrak{p}q_1 & q_{2}^\dagger\mathfrak{p}q_2 & q_{2}^\dagger\mathfrak{p}q_3 \\
q_{3}^\dagger\mathfrak{p}1 & q_{3}^\dagger\mathfrak{p}q_{23} & q_{3}^\dagger\mathfrak{p}q_{31} & q_{3}^\dagger\mathfrak{p}q_{12} & q_{3}^\dagger\mathfrak{p}q_{321} & q_{3}^\dagger\mathfrak{p}q_1 & q_{3}^\dagger\mathfrak{p}q_2 & q_{3}^\dagger\mathfrak{p}q_3
\end{smallmatrix}\right)=\left(\begin{smallmatrix}
\nu_{R_1}&u_{R_1}^r&u_{R_1}^y&u_{R_1}^b&\bar{e}_{L1}&\bar{d}_{L1}^{\bar{r}}&\bar{d}_{L1}^{\bar{y}}&\bar{d}_{L1}^{\bar{b}} \\
\nu_{R_2}&u_{R_2}^r&u_{R_2}^y&u_{R_2}^b&\bar{e}_{L2}&\bar{d}_{L2}^{\bar{r}}&\bar{d}_{L2}^{\bar{y}}&\bar{d}_{L2}^{\bar{b}} \\
\nu_{L_1}&u_{L_1}^r&u_{L_1}^y&u_{L_1}^b&\bar{e}_{R1}&\bar{d}_{R1}^{\bar{r}}&\bar{d}_{R1}^{\bar{y}}&\bar{d}_{R1}^{\bar{b}} \\
\nu_{L_2}&u_{L_2}^r&u_{L_2}^y&u_{L_2}^b&\bar{e}_{R2}&\bar{d}_{R2}^{\bar{r}}&\bar{d}_{R2}^{\bar{y}}&\bar{d}_{R2}^{\bar{b}}  \\
e_{L1}&d_{L1}^r&d_{L1}^y&d_{L1}^b&\bar{\nu}_{R1}&\bar{u}_{R1}^{\bar{r}}&\bar{u}_{R1}^{\bar{y}}&\bar{u}_{R1}^{\bar{b}} \\
e_{L2}&d_{L2}^r&d_{L2}^y&d_{L2}^b&\bar{\nu}_{R2}&\bar{u}_{R2}^{\bar{r}}&\bar{u}_{R2}^{\bar{y}}&\bar{u}_{R2}^{\bar{b}} \\
e_{R1}&d_{R1}^r&d_{R1}^y&d_{R1}^b&\bar{\nu}_{L1}&\bar{u}_{L1}^{\bar{r}}&\bar{u}_{L1}^{\bar{y}}&\bar{u}_{L1}^{\bar{b}} \\
e_{R2}&d_{R2}^r&d_{R2}^y&d_{R2}^b&\bar{\nu}_{L2}&\bar{u}_{L2}^{\bar{r}}&\bar{u}_{L2}^{\bar{y}}&\bar{u}_{L2}^{\bar{b}} \\
\end{smallmatrix}\right)
\end{equation}
where the complex matrix elements of the right-hand side are suggestively labeled to make obvious association with elementary fermion states. Here, the superscripts refer to color, whereas the subscripts denote chiral indices. Notice also that the even grade elements (composing spinors) compose the two diagonal $4\times4$ blocks and the odd grade elements compose the two off-diagonal $4\times4$ blocks. Writing out the third minimal left ideal explicitly, for example, yields:
\begin{eqnarray}
\nonumber P_3&=&u_{R_1}^y\mathfrak{p}q_{31}+u_{R_2}^yq_{23}^\dagger\mathfrak{p}q_{31}+u_{L_1}^yq_{31}^\dagger\mathfrak{p}q_{31}+u_{L_2}^yq_{12}^\dagger\mathfrak{p}q_{31}\\
&+&d_{L1}^yq_{321}^\dagger\mathfrak{p}q_{31}+d_{L2}^yq_{1}^\dagger\mathfrak{p}q_{31}+d_{R1}^yq_{2}^\dagger\mathfrak{p}q_{31}+d_{R2}^yq_{3}^\dagger\mathfrak{p}q_{31},
\end{eqnarray}
where $u_{R_1}^y\in\mathbb{C}$ e.t.c. are complex numbers and $\mathfrak{p}q_{31}$ e.t.c. are elements of $P_3\in C\ell(6)$. The identification of the basis states of the ideals with specific quark and lepton states in determined from the left and right adjoint actions of $C\ell(6)$ on the ideals. 
\subsection{The left adjoint actions $C\ell(4)_{\mathrm{Dirac}}\otimes_{\mathbb{C}}C\ell(2)_{\mathrm{isospin}}$}
Take a standard orthonormal basis $\{e_i\}$, $i=1,\cdots,2n$ for $C\ell(2n)$, together with a matrix representation $\{\rho(e_i)\}$ of $\mathbb{C}(2^n)\cong C\ell(2n)$. To extend to a matrix representation $\mathbb{C}(2^{n+1})$ of $C\ell(2n+2)$ we can set\footnote{where $i=1,\cdots,2n$ and the $\rho(e_i)$ inside the matrices are the matrix representations of the smaller Clifford algebra.}:
\begin{equation}
\rho(e_i)\longrightarrow \rho(e_i)\equiv\left(\begin{smallmatrix}
\rho(e_i)&0 \\
0&-\rho(e_i)
\end{smallmatrix}\right),\quad \rho(e_{2n+1})\equiv\left(\begin{smallmatrix}
0&1 \\
1&0
\end{smallmatrix}\right),\quad \rho(e_{2n+2})\equiv\left(\begin{smallmatrix}
0&-i \\
i&0
\end{smallmatrix}\right).
\end{equation}
Note that $\rho(e_{2n+1})$ and $\rho(e_{2n+2})$ are $su(2)$ generators, with the third ``$z$-component'' $su(2)$ generator given automatically by $-i\rho(e_{2n+1})\rho(e_{2n+2})$. The $su(2)$ generators can therefore be used to generate a $C\ell(2)$ algebra. It follows that extending the Dirac algebra $C\ell(4)_{\mathrm{Dirac}}$, generated by $\Gamma_{\mu}$, via the inclusion of isospin $T_i\in su(2)_L$ generators,  is equivalent to extending $C\ell(4)_{\mathrm{Dirac}}$ to $C\ell(6)_{\mathrm{Dirac+isospin}}$ via the tensor product $C\ell(4)_{\mathrm{Dirac}}\otimes_{\mathbb{C}}C\ell(2)_{\mathrm{isospin}}$\footnote{More details on the extension from the $C\ell(4)$ Dirac algebra to $C\ell(6)$ via the inclusion of weak symmetry can be found in appendix B of \cite{Stoica2018}.}. Explicitly, the matrix representations for $\Gamma_{\mu}$ and $T_i$ are given by
\begin{eqnarray}
\rho(\Gamma_{0})=\left(\begin{smallmatrix}
0&1_2 &0 & 0 \\
1_2&0&0 &0\\
0 &0 &0 &1_2\\
0 &0 &1_2 & 0
\end{smallmatrix}\right),\qquad \rho(\Gamma_{i})=\left(\begin{smallmatrix}
0&\sigma_i &0 & 0 \\
-\sigma_i &0&0 &0\\
0 &0 &0 &-\sigma_i\\
0 &0 &\sigma_i & 0
\end{smallmatrix}\right),\qquad i=1,2,3,
\end{eqnarray}
where $\sigma_i$ are the usual Pauli matrices\footnote{$\sigma_1=\left(\begin{smallmatrix}
0&1 \\
1&0
\end{smallmatrix}\right),\quad \sigma_2=\left(\begin{smallmatrix}
0&-i \\
i&0
\end{smallmatrix}\right),\quad \sigma_1=\left(\begin{smallmatrix}
1&0 \\
0&-1
\end{smallmatrix}\right).
$}
, and
\begin{eqnarray}
\rho(T_1)=2i\left(\begin{smallmatrix}
0& 0 &0 & 0 \\
0&0&1_2 &0\\
0 &1_2 &0 &0\\
0 &0 &0 & 0
\end{smallmatrix}\right),\quad \rho(T_2)=2\left(\begin{smallmatrix}
0& 0 &0 & 0 \\
0&0&-1_2 &0\\
0 &1_2 &0 &0\\
0 &0 &0 & 0
\end{smallmatrix}\right),\quad \rho(T_3)=2i\left(\begin{smallmatrix}
0& 0 &0 & 0 \\
0&-1_2&0 &0\\
0 &0 &1_2 &0\\
0 &0 &0 & 0
\end{smallmatrix}\right).
\end{eqnarray}

Consider now the left action:
\begin{equation}
\mathbb{C}(8)\triangleright \rho(P_i),\quad i=1,2,3,4,5,6,7,8.
\end{equation}
Multiplying a minimal left ideal matrix representation from the left results in rows being interchanged, but not columns. Hence left multiplications constitute transformations within an ideal.
The left multiplication then allows for the identification of basis states of an individual minimal left ideal with two sets of Lorentzian indices $\{\ell_1,\ell_2,\ell_3,\ell_4\}$ that each get further sub-divided into right chiral indices $\{R_1,R_2\}$ and left chiral indices $\{L_1,L_2\}$.
\subsection{The right adjoint actions $U(3)_{\mathrm{ec}}$}
Next, consider the right action:
\begin{equation}
\rho(P_i)\triangleleft \mathbb{C}(8)\quad i=1,2,3,4,5,6,7,8.
\end{equation}
When a minimal left ideal matrix representation is right multiplied by a square matrix, the columns are interchanged rather than the rows. The right adjoint action algebra $C\ell(6)$, having a maximal totally isotropic subspace (MTIS) symmetry group $U(3)$, transforms the columns of $\mathbb{C}(8)$ so that $\rho(P_1)$ transforms as a color singlet, $\{\rho(P_2),\rho(P_3),\rho(P_4)\}$ transform as a color triplet, $\rho(P_5)$ transforms as an color anti-singlet and the remaining minimal left ideals $\{\rho(P_6),\rho(P_7),\rho(P_8)\}$ transform as a color anti-triplet. This corresponds to the color assignments $r,y,b$ and $\bar{r},\bar{y}, \bar{b}$. The $U(1)$ electric charge operator
acts from both left and right, unlike the $SU(3)$ operators. The generators $\{\rho(Q),\rho(\Lambda_i)\}$, $i=1,\cdots,8$ of the electrocolor MTIS symmetry group $U(3)_{\mathrm{ec}}$ are constructed from the matrices that are right multiplied onto the minimal left ideals $\rho(P_i)$.  We write in symbolic form:
\begin{equation}
C\ell(6)_{\mathrm{Dirac+isospin}}\cong\mathbb{C}(8)\triangleright \{\rho(P_1),\cdots,\rho(P_8)\}\triangleleft \mathbb{C}(8)\cong C\ell(6)\supset U(3)_{\mathrm{ec}}.
\end{equation}
Therefore, what \cite{Stoica2018} shows is that one generation of leptons and quarks with the desired symmetries is contained in $C\ell(6)$\footnote{A nice zeroth order prediction of the Weinberg mixing angle $\theta_W$ is also there obtained.}. Finally, it should be noted that the action of the $C\ell(6)_{\textrm{Dirac+isospin}}$ algebra on the ideals is independent of the colors and the action of $U(3)_{ec}$.
\section{From $C\ell(6)$ to $C\ell(8)$ via the inclusion of spin}
\label{sec:DWIS}
Here we show that including spin into the $C\ell(6)$ results reviewed in the previous section enlarges the algebra of left adjoint actions to $C\ell(8)\cong\mathbb{C}(16)$. Similarly, in this framework the minimal left ideals are now enlarged to be column matrices (pinors) of $\mathbb{C}(16)$. There are now 16 minimal left ideals $P_i$. We will then consider the left action of $C\ell(8)\cong\mathbb{C}(16)$ on these minimal left ideals.

Starting with $C\ell(6)\cong C\ell(4)_{\mathrm{Dirac}}\otimes_{\mathbb{C}} C\ell(2)_{\mathrm{isospin}}$, we now wish to include the generators $J_i\in su(2)_{\mathrm{spin}}$. As with $C\ell(2)_{\mathrm{isospin}}$ generated from the generators of $su(2)_L$, so the generators of $su(2)_{\mathrm{spin}}$ generate a second factor of $C\ell(2)$, which we label $C\ell(2)_{\mathrm{spin}}$. Taken altogether, this gives $C\ell(8)$, namely:
\begin{equation}
C\ell(4)_{\mathrm{Dirac}}\otimes_{\mathbb{C}} C\ell(2)_{\mathrm{isospin}}\otimes_{\mathbb{C}} C\ell(2)_{\mathrm{spin}}\cong C\ell(8)_{\mathrm{Dirac+isospin+spin}}.
\end{equation}

The rows of our representation can be chosen so that the matrix representations of $C\ell(6)_{\mathrm{Dirac+isospin}}\subset C\ell(8)_{\mathrm{Dirac+isospin+spin}}$ found in \cite{Stoica2018} extend here via a simple direct sum:
\begin{equation}
\rho(\Gamma_{\mu})\mapsto \rho(\Gamma_{\mu})\oplus \rho(\Gamma_{\mu}),\quad \rho(T_i)\mapsto\rho(T_i)\oplus\rho(T_i).
\end{equation}
In this case the action of $C\ell(6)_{\mathrm{Dirac+isospin}}\subset C\ell(8)_{\mathrm{Dirac+isospin+spin}}$ on the minimal left ideals act irreducibly on the top eight rows of $\mathbb{C}(16)$, and the bottom eight rows of $\mathbb{C}(16)$. For this choice of matrix representations, the spin generators are given in terms of $8\times 8$ blocks by:
\begin{equation}
\nonumber\rho(J_1)=\left(\begin{smallmatrix}
0&1 \\
1&0
\end{smallmatrix}\right),\quad \rho(J_2)=\left(\begin{smallmatrix}
0&-i \\
i&0
\end{smallmatrix}\right),\quad \rho(J_3)=\left(\begin{smallmatrix}
1&0 \\
0&-1
\end{smallmatrix}\right),\quad \rho(J_+)=\left(\begin{smallmatrix}
0&1 \\
0&0
\end{smallmatrix}\right),\quad \rho(J_-)=\left(\begin{smallmatrix}
0&0 \\
1&0
\end{smallmatrix}\right).
\end{equation}
This will prove useful when it comes to considering the left action of $C\ell(8)\cong\mathbb{C}(16)$ on the minimal left ideals, which determines the transformation properties of each element of the minimal left ideals. 

Physical quantum states correspond to spinors rather than pinors. The $C\ell(8)$ spinors are elements of the even Clifford grade part of the algebra: $C\ell^+(8)\cong C\ell(7)\cong\mathbb{C}(8)\oplus\mathbb{C}(8)\subset\mathbb{C}(16)$\footnote{The algebra $C\ell(7)$ in relation to a geometric approach to the SM was also considered in \cite{trayling2004cl}. There, the extra four spacelike dimensions are the minimal required to incorporate all the fermions of one generation into a single spinor. Interestingly, in that work these extra four dimensions for a basis form the Higgs isodoublet field.}. This means that the physically allowed transformations are exclusively those that map spinors to spinors. Transformations that map between pinors and spinors are excluded. The full $C\ell(8)$ mathematical left adjoint actions can be symbolically given as
\begin{equation}
C\ell(8)_{\mathrm{Dirac+isospin+spin}}\triangleright \{P_1,\cdots,P_{16}\}.
\end{equation}
However, $C\ell(8)$ includes the unphysical transformations that map between spinor and pinor spaces.

The left-action weak transformations preserve the spinor spaces and so those transformations are physical, and allow for the existence of the weak $W^\pm$ and $Z^0$ vector bosons. The spin ladder operators an the other hand do not preserve the spinor spaces and therefor do not correspond to physically allowed transformations. No vector bosons associated with $J_\pm$ transformations are observed in nature. For example, consider
\begin{eqnarray}
\rho(J_+)\left(\begin{matrix}
\mathbb{C}(8) &0 \\
0& \mathbb{C}(8)
\end{matrix}\right)=\left(\begin{matrix}
0 &\mathbb{C}(8) \\
0& 0
\end{matrix}\right).
\end{eqnarray}
The maximal left action subalgebra of $C\ell(8)_{\mathrm{Dirac+isospin+spin}}$ that preserves the space of spinors is therefore $C\ell(7)_{\textrm{Dirac+isospin+spin-z}}\subset C\ell(8)_{\mathrm{Dirac+isospin+spin}}$ where the subscript ``spin-$z$'' is meant to indicate that the diagonal spinor space preserving $J_3$ generator is kept while the non-spinor-space-preserving  $J_\pm$ are not kept\footnote{This raises the question of the possible existence of a ``spin-$z$ vector boson''. We have identified the $z$-component as a Clifford bivector and the other two components as vectors. The same is true in the weak isospin case. In the weak isospin case the $W^\pm$ bosons exist. Interestingly the physical $Z^0$ boson does not follow directly from $T_z$, but mixes with a $U(1)_Y$ hypercharge. $T_z$ does not by itself result in a boson.It could be that there is nothing for $J_z$ to mix with in order to produce a physical boson.}.
\section{$C\ell(8)$ and the MTIS symmetry group $U(4)$}
\label{sec:MTIS}
Having considered the left adjoint actions in the previous section, we now focus on the right adjoint actions. We start by writing a Witt basis for $C\ell(8)$ in terms of the generators $\{q_i,q_i^\dagger\}$, $i=1,2,3,4$ which satisfy the algebraic properties
\begin{equation}
q_i^2=(q_i^\dagger)^2=\{q_i,q_j\}=\{q_i^\dagger,q_j^\dagger\}=0,\quad \{q_i,q_j^\dagger\}=\delta_{ij},\quad i,j=1,2,3,4.
\end{equation}
The minimal left ideals $P_i$, $i=1,\cdots,16$ of $C\ell(8)$ can be expressed in terms of column vectors $\rho(P_i)\in\mathbb{C}(16)\cong C\ell(8)$ as a $16\times16$ matrix $(R_1,R_2)$ where $R_1$ and $R_2$ are two $16\times 8$ matrices displayed symbolically in Appendix A. As was the case with $\mathbb{C}(8)$, we again see that for $\mathbb{C}(16)$, the two diagonal $8\times 8$ blocks correspond to the spinors. Here $p\equiv q_{1234}q_{4321}^\dagger\equiv q_1q_2q_3q_4q_4^\dagger q_3^\dagger q_2^\dagger q_1^\dagger\in C\ell^+(8)$, the even subalgebra of $C\ell(8)$. We determine the particle identifications for each element by considering the left action of $C\ell(8)_L$ on the minimal left ideals and the right action of the MTIS symmetry group generators. To find the MTIS symmetry generators we set
\begin{equation}
\boldsymbol{\alpha}=\sum_{i=1}^4c_iq_i,\quad \boldsymbol{\alpha}'=\sum_{i=1}^4c_i'q_i,\quad c_i,c_i'\in\mathbb{C}
\end{equation}
and calculate the most general hermitian operator $\mathcal{H}=\boldsymbol{\alpha}^{'\dagger}\boldsymbol{\alpha}+\boldsymbol{\alpha}^\dagger\boldsymbol{\alpha}'$. The result is sixteen generators of the Lie algebra $u(4)$. The MTIS $U(4)$ generators are:
\begin{eqnarray}
\nonumber\Lambda_1&=&-(q_1^\dagger q_2+q_2^\dagger q_1),\quad \Lambda_2=i(q_2^\dagger q_1-q_1^\dagger q_2), \\
\nonumber\Lambda_4&=&-(q_3^\dagger q_1+q_1^\dagger q_3)\quad \Lambda_5=i(q_3^\dagger q_1-q_1^\dagger q_3), \\
\nonumber\Lambda_6&=&-(q_2^\dagger q_3+q_3^\dagger q_2)\quad \Lambda_7=i(q_3^\dagger q_2-q_2^\dagger q_3), \\
\nonumber\Lambda_3&=&q_2^\dagger q_2-q_1^\dagger q_1,\quad \Lambda_8=\frac{-1}{\sqrt{3}}(q_1^\dagger q_1+q_2^\dagger q_2-2q_3^\dagger q_3), \\
\Lambda_9&=&-(q_1^\dagger q_4+q_4^\dagger q_1),\quad\Lambda_{10}=i(q_1^\dagger q_4-q_4^\dagger q_1), \\
\nonumber\Lambda_{11}&=&-(q_2^\dagger q_4+q_4^\dagger q_2)\quad \Lambda_{12}=i(q_2^\dagger q_4-q_4^\dagger q_2), \\
\nonumber\Lambda_{13}&=&-(q_3^\dagger q_4+q_4^\dagger q_3)\quad \Lambda_{14}=i(q_3^\dagger q_4-q_4^\dagger q_3). \\
\nonumber\Lambda_{15}&\propto&f_1q_1^\dagger q_1+f_2 q_2^\dagger q_2+f_3q_3^\dagger q_3+f_4q_4^\dagger q_4,\quad f_i\in\mathbb{R} \\
\nonumber Q &=&\frac{1}{3}\sum_{i=1}^4q_i^\dagger q_i.
\end{eqnarray} 
The color subgroup $SU(3)_{\mathrm{c}}\subset U(4)$ is generated by $\Lambda_1,\cdots,\Lambda_8$. It is readily checked that $q_4,q_4^{\dagger}$ commute with this $SU(3)$ subgroup. Hence when it comes to identifying the color quantum numbers associated with each element of each minimal left ideal, one can ignore the $q_4$'s and $q_4^\dagger$'s that are present both explicitly, and also inside the ``vacuum'' idempotent $p=q_{1234}q_{1234}^\dagger$. For example, suppose we wanted to know how the state $q_1^\dagger q_4^\dagger pq_{32}$ transforms under $SU(3)_{\mathrm{c}}$. We simply calculate $[\Lambda_i,q_{14}^\dagger pq_{32}]$ for $i=1,\cdots,8$. This state turns out to be color red. To see how this works we illustrate with $\Lambda_1$:
\begin{eqnarray}
[\Lambda_1,q_{14}^\dagger pq_{32}]&=&0+q_{14}^\dagger q_{1234}q_{4321}^\dagger q_{32}(q_1^\dagger q_2+q_2^\dagger q_1)\\
\nonumber &=&-q_{14}^\dagger q_{1234}q_{4321}^\dagger q_3q_1^\dagger(q_2)^2+q_{14}^\dagger q_{1234}q_{4321}^\dagger q_3(-q_2^\dagger q_2+1)q_1 \\
\nonumber&=&0+q_{14}^\dagger q_{1234}q_{4321}^\dagger q_3(-q_2^\dagger q_2+1)q_1 \\
\nonumber&=&q_{14}^\dagger q_{1234}q_{43}^\dagger(q_2^\dagger)^2q_{321}+q_{14}^\dagger q_{1234}q_{4321}^\dagger q_{31} \\
\nonumber&=&0+q_{14}^\dagger q_{1234}q_{4321}^\dagger q_{31} \\
\nonumber&=&q_{14}^\dagger p q_{31}.
\end{eqnarray}
Each of the sixteen minimal left ideals of $C\ell(8)$ is of constant color. By obvious color labellings the color assignments of the minimal left ideals $P_i\in C\ell(8)$, $i=1,\cdots,16$ may be written as:
\begin{equation}
P_1,P_2^r,P_3^y,P_4^b,P_5,P_6^{\bar{r}},P_7^{\bar{y}},P_8^{\bar{b}},P_9,P_{10}^{r},P_{11}^{y},P_{12}^{b},P_{13},P_{14}^{\bar{r}},P_{15}^{\bar{y}},P_{16}^{\bar{b}},
\end{equation}
where a lack of superscript means that the ideal transforms as a color (anti)singlet. With the choice of matrix representation given by the column arrangement as in $(R_1,R_2)$, the matrix representations are a simple direct sum extension to that found in \cite{Stoica2018}  :
\begin{equation}
\rho(\Lambda_i)\mapsto\rho(\Lambda_i)\oplus\rho(\Lambda_i),\quad i=1,\cdots,8.
\end{equation}
The non-electrocolor $U(4)$ transformations, generated by $\Lambda_9$-$\Lambda_{15}$, do not map spinors to spinors, and are therefore unphysical. The electrocolor subgroup $U(3)_{\mathrm{ec}}\subset U(4)$ of $U(4)$ preserves the spinor spaces so those transformations are physical, and in fact, the electrocolor MTIS subgroup is the maximal subgroup that does so\footnote{The works \cite{dahm2012microscopic,dahm2010symmetry} look at $SU^*(4)$, and a symmetry reduction scheme of the Dirac algebra. In that context the QCD $SU(3)$ group likewise arises as part of $SU(4)$.}. 
\section{Spinors and triality: three generations}
\label{sec:3gens}
In this section we demonstrate that the triality automorphism of $\textrm{Spin}(8)$ extends all the results from a single generation to exactly three generations. This is the second main result of this paper.

The basis states of the sixteen minimal left ideals $P_1,\cdots,P_{16}$ are elements of the Clifford-Lipschitz group $P_i\in\Gamma(8,\mathbb{C})\subset C\ell(8)$ which is the group generated by all invertible $s\in C\ell(8)$:
\begin{equation}
\Gamma(8,\mathbb{C})\equiv\{s\in C\ell^+(8)\cup C\ell^-(8)|\forall x\in\mathbb{C}^{8},\quad sxs^{-1}\in\mathbb{C}^{8}\}.
\end{equation}
The normalized subgroup of the Clifford-Lipschitz group $\Gamma(8,\mathbb{C})$ is the Pin group:
\begin{equation}
\mathrm{Pin}(8,\mathbb{C})\equiv\{s\in\Gamma(8,\mathbb{C})|s\tilde{s}=\pm1\}
\end{equation}
where, for a Clifford algebra $C\ell(p,q)$ with orthonormal basis $e_i$, $i=1....p+q$,
\begin{equation}
\tilde{}:e_{i_1}e_{i_2}\cdots e_{i_m}\mapsto e_{i_m}\cdots e_{i_2}e_{i_1}
\end{equation}
is Clifford reversion and $\widetilde{uv}=\tilde{v}\tilde{u}$\footnote{The Clifford inverse is given by $u^{-1}=\frac{\bar{u}}{u\bar{u}}$, where $\bar{u}$ is the Clifford conjugate defined as the composition of Clifford reversion and Clifford grade involution $\hat{u}$, defined as
\begin{equation}
\hat{}:\left\langle u\right\rangle_{\mathrm{even}}\rightarrow +\left\langle u\right\rangle_{\mathrm{even}},\quad \hat{}:\left\langle u\right\rangle_{\mathrm{odd}}\rightarrow -\left\langle u\right\rangle_{\mathrm{odd}}.
\end{equation} 
}. The sixteen minimal left ideals $P_i$ are pinors. The even subgroup of the Pin group is the Spin group:
\begin{equation}
\mathrm{Spin}(8,\mathbb{C})\equiv\mathrm{Pin(8,\mathbb{C})}\cap C\ell^+(8)=\{s\in C\ell^+(8)|s\tilde{s}=\pm1\}.
\end{equation}
To be consistent with probability conservation in quantum mechanics we require that $s\tilde{s}=+1$, 
\begin{equation}
\mathrm{Spin}_+(8,\mathbb{C})\equiv\{s\in C\ell^+(8)|s\tilde{s}=1\}.
\end{equation}
Physical states should then correspond to spinors in $\mathrm{Spin}_+(8,\mathbb{C})$. Note that $C\ell^+(8)\cong C\ell(7)\cong\mathbb{C}(8)\oplus\mathbb{C}(8)$. Our choice of matrix representations of $R_1$ and $R_2$ makes this isomorphism easy to picture. 

Triality is an automorphism of Spin$(8)$. That is, it is an automorphism of spinors rather than pinors. We therefore restrict our attention to the spinors, which are the minimal left ideals $S_i$ restricted by setting the coefficients of the odd grade Clifford elements in the pinors $P_i$ equal to zero. The $C\ell(8)$ spinors are given symbolically by the columns of $(R_{1,s},R_{2,s})$. $R_{1,s}$ is obtained from $R_1$ by simply setting the last eight rows (corresponding to odd grade elements) equal to zero. Similarly, $R_{2,s}$, is obtained by setting the first eight rows in $R_2$ equal to zero. $R_{1,s}$ and $R_{2,s}$ are shown symbolically in Appendix A.
\subsection{The action of triality on spinors}
Triality is a non-linear automorphism of order three. We are interested in the action of triality on the spinors of $C\ell(8)$. The two spinor representations of $\mathrm{Spin}(8)$, $S_8^+$ and $S_8^-$, as well as the fundamental vector representation $V_8$ are each eight-dimensional. The action of the triality map on these representations is given by Trial$:\{V_8,S_8^+,S_8^-\}\rightarrow \{S_8^+,S_8^-,V_8\}$.

For three $8\times 8$ matrices $A,B,C$, given respectively by:
\begin{equation}
A=\left(\begin{smallmatrix}
\nu_{e,R_1}&u_{R_1}^{r}&u_{R_1}^{y}&u_{R_1}^{b}&e^{+}_{L1}&\bar{d}_{L1}^{\bar{r}}&\bar{d}_{L1}^{\bar{y}}&\bar{d}_{L1}^{\bar{b}} \\
\nu_{e,R_2}&u_{R_2}^{r}&u_{R_2}^{y}&u_{R_2}^{b}&e^{+}_{L2}&\bar{d}_{L2}^{\bar{r}}&\bar{d}_{L2}^{\bar{y}}&\bar{d}_{L2}^{\bar{b}} \\
\nu_{e,L_1}&u_{R_1}^{r}&u_{L_1}^{y}&u_{L_1}^{b}&e^{+}_{R1}&\bar{d}_{R1}^{\bar{r}}&\bar{d}_{R1}^{\bar{y}}&\bar{d}_{R1}^{\bar{b}} \\
\nu_{e,L_2}&u_{R_2}^{r}&u_{L_2}^{y}&u_{L_2}^{b}&e^{+}_{R2}&\bar{d}_{R2}^{\bar{r}}&\bar{d}_{R2}^{\bar{y}}&\bar{d}_{R2}^{\bar{b}} \\
e^{-}_{L1}&d_{L1}^{r}&d_{L1}^{y}&d_{L1}^{b}&\bar{\nu}_{e,R1}&\bar{u}_{R1}^{\bar{r}}&\bar{u}_{R1}^{\bar{y}}&\bar{u}_{R1}^{\bar{b}} \\
e^{-}_{L2}&d_{L2}^{r}&d_{L2}^{y}&d_{L2}^{b}&\bar{\nu}_{e,R2}&\bar{u}_{R2}^{\bar{r}}&\bar{u}_{R2}^{\bar{y}}&\bar{u}_{R2}^{\bar{b}} \\
e^{-}_{R1}&d_{R1}^{r}&d_{R1}^{y}&d_{R1}^{b}&\bar{\nu}_{e,L1}&\bar{u}_{L1}^{\bar{r}}&\bar{u}_{L1}^{\bar{y}}&\bar{u}_{L1}^{\bar{b}} \\
e^{-}_{R2}&d_{R2}^{r}&d_{R2}^{y}&d_{R2}^{b}&\bar{\nu}_{e,L2}&\bar{u}_{L2}^{\bar{r}}&\bar{u}_{L2}^{\bar{y}}&\bar{u}_{L2}^{\bar{b}} \\
\end{smallmatrix}\right),
\end{equation}
\begin{equation}
B=\left(\begin{smallmatrix}
\nu_{\tau,R_1}&t_{R_1}^{r}&t_{R_1}^{y}&t_{R_1}^{b}&\tau^{+}_{L1}&\bar{b}_{L1}^{\bar{r}}&\bar{b}_{L1}^{\bar{y}}&\bar{b}_{L1}^{\bar{b}} \\
\nu_{\tau,R_2}&t_{R_2}^{r}&t_{R_2}^{y}&t_{R_2}^{b}&\tau^{+}_{L2}&\bar{b}_{L2}^{\bar{r}}&\bar{b}_{L2}^{\bar{y}}&\bar{b}_{L2}^{\bar{b}} \\
\nu_{\tau,L_1}&t_{R_1}^{r}&t_{L_1}^{y}&t_{L_1}^{b}&\tau^{+}_{R1}&\bar{b}_{R1}^{\bar{r}}&\bar{b}_{R1}^{\bar{y}}&\bar{b}_{R1}^{\bar{b}} \\
\nu_{\tau,L_2}&t_{R_2}^{r}&t_{L_2}^{y}&t_{L_2}^{b}&\tau^{+}_{R2}&\bar{b}_{R2}^{\bar{r}}&\bar{b}_{R2}^{\bar{y}}&\bar{b}_{R2}^{\bar{b}}  \\
\tau^{-}_{L1}&b_{L1}^{r}&b_{L1}^{y}&b_{L1}^{b}&\bar{\nu}_{\tau,R1}&\bar{t}_{R1}^{\bar{r}}&\bar{t}_{R1}^{\bar{y}}&\bar{t}_{R1}^{\bar{b}} \\
\tau^{-}_{L2}&b_{L2}^{r}&b_{L2}^{y}&b_{L2}^{b}&\bar{\nu}_{\tau,R2}&\bar{t}_{R2}^{\bar{r}}&\bar{t}_{R2}^{\bar{y}}&\bar{t}_{R2}^{\bar{b}} \\
\tau^{-}_{R1}&b_{R1}^{r}&b_{R1}^{y}&b_{R1}^{b}&\bar{\nu}_{\tau,L1}&\bar{t}_{L1}^{\bar{r}}&\bar{t}_{L1}^{\bar{y}}&\bar{t}_{L1}^{\bar{b}} \\
\tau^{-}_{R2}&b_{R2}^{r}&b_{R2}^{y}&b_{R2}^{b}&\bar{\nu}_{\tau,L2}&\bar{t}_{L2}^{\bar{r}}&\bar{t}_{L2}^{\bar{y}}&\bar{t}_{L2}^{\bar{b}} \\
\end{smallmatrix}\right),
\end{equation}
\begin{equation}
C=\left(\begin{smallmatrix}
\nu_{\mu,R_1}&c_{R_1}^{r}&c_{R_1}^{y}&c_{R_1}^{b}&\mu^{+}_{L1}&\bar{s}_{L1}^{\bar{r}}&\bar{s}_{L1}^{\bar{y}}&\bar{s}_{L1}^{\bar{b}}\\
\nu_{\mu,R_2}&c_{R_2}^{r}&c_{R_2}^{y}&c_{R_2}^{b}&\mu^{+}_{L2}&\bar{s}_{L2}^{\bar{r}}&\bar{s}_{L2}^{\bar{y}}&\bar{s}_{L2}^{\bar{b}} \\
\nu_{\mu,L_1}&c_{R_1}^{r}&c_{L_1}^{y}&c_{L_1}^{b}&\mu^{+}_{R1}&\bar{s}_{R1}^{\bar{r}}&\bar{s}_{R1}^{\bar{y}}&\bar{s}_{R1}^{\bar{b}} \\
\nu_{\mu,L_2}&c_{R_2}^{r}&c_{L_2}^{y}&c_{L_2}^{b}&\mu^{+}_{R2}&\bar{s}_{R2}^{\bar{r}}&\bar{s}_{R2}^{\bar{y}}&\bar{s}_{R2}^{\bar{b}} \\
\mu^{-}_{L1}&s_{L1}^{r}&s_{L1}^{y}&s_{L1}^{b}&\bar{\nu}_{\mu,R1}&\bar{c}_{R1}^{\bar{r}}&\bar{c}_{R1}^{\bar{y}}&\bar{c}_{R1}^{\bar{b}} \\
\mu^{-}_{L2}&s_{L2}^{r}&s_{L2}^{y}&s_{L2}^{b}&\bar{\nu}_{\mu,R2}&\bar{c}_{R2}^{\bar{r}}&\bar{c}_{R2}^{\bar{y}}&\bar{c}_{R2}^{\bar{b}} \\
\mu^{-}_{R1}&s_{R1}^{r}&s_{R1}^{y}&s_{R1}^{b}&\bar{\nu}_{\mu,L1}&\bar{c}_{L1}^{\bar{r}}&\bar{c}_{L1}^{\bar{y}}&\bar{c}_{L1}^{\bar{b}}\\
\mu^{-}_{R2}&s_{R2}^{r}&s_{R2}^{y}&s_{R2}^{b}&\bar{\nu}_{\mu,L2}&\bar{c}_{L2}^{\bar{r}}&\bar{c}_{L2}^{\bar{y}}&\bar{c}_{L2}^{\bar{b}} \\
\end{smallmatrix}\right),
\end{equation}
the action of triality is as follows \cite{lounesto2001caa}:
\begin{equation}
\mathrm{Trial}\left(\begin{smallmatrix}
A&0 \\
0&B
\end{smallmatrix}\right)=\left(\begin{smallmatrix}
C&0 \\
0&A
\end{smallmatrix}\right),\quad \mathrm{Trial}\left(\begin{smallmatrix}
C&0 \\
0&A
\end{smallmatrix}\right)=\left(\begin{smallmatrix}
B&0 \\
0&C
\end{smallmatrix}\right),\quad\mathrm{Trial}\left(\begin{smallmatrix}
B&0 \\
0&C
\end{smallmatrix}\right)=\left(\begin{smallmatrix}
A&0 \\
0&B
\end{smallmatrix}\right).
\end{equation}
The action of triality on each diagonal $\mathbb{C}(8)$ block matrix is to permute through the different diagonal $\mathbb{C}(8)$ block matrices.
We may further consider the action of $\rho(J_{3})$ on these matrices\footnote{The action is simply left matrix multiplication by $\rho(J_3)$.}. We find that the top $8\times8$ blocks are spin-up and the bottom blocks are spin-down, which will be accounted for by introducing obvious notation with up and down arrows:
\begin{equation}
\mathrm{Trial}\left(\begin{smallmatrix}
A^\uparrow&0 \\
0&B^\downarrow
\end{smallmatrix}\right)=\left(\begin{smallmatrix}
C^\uparrow&0 \\
0&A^\downarrow
\end{smallmatrix}\right),\quad \mathrm{Trial}\left(\begin{smallmatrix}
C^\uparrow&0 \\
0&A^\downarrow
\end{smallmatrix}\right)=\left(\begin{smallmatrix}
B^\uparrow&0 \\
0&C^\downarrow
\end{smallmatrix}\right),\quad\mathrm{Trial}\left(\begin{smallmatrix}
B^\uparrow&0 \\
0&C^\downarrow
\end{smallmatrix}\right)=\left(\begin{smallmatrix}
A^\uparrow&0 \\
0&B^\downarrow
\end{smallmatrix}\right).
\end{equation}
Next consider the respective actions of $\rho(J_+)\circ\mathrm{Trial}$ and $\rho(J_-)\circ\mathrm{Trial}$. For example,
\begin{equation}
\rho(J_+)\circ\mathrm{Trial}\left(\begin{smallmatrix}
A^\uparrow&0 \\
0&B^\downarrow
\end{smallmatrix}\right)=\left(\begin{smallmatrix}
0&A^\uparrow \\
0&0
\end{smallmatrix}\right),\quad \rho(J_-)\circ\mathrm{Trial}\left(\begin{smallmatrix}
A^\uparrow&0 \\
0&B^\downarrow
\end{smallmatrix}\right)=\left(\begin{smallmatrix}
0&0 \\
C^{\downarrow}&0
\end{smallmatrix}\right).
\end{equation}
We can now examine the left actions and right actions of $C\ell(8)$ on the three matrices $X=A^{\uparrow}\oplus B^{\downarrow},\mathrm{Trial}(X)=C^{\uparrow}\oplus A^{\downarrow},\mathrm{Trial}\circ\mathrm{Trial}(X)=B^{\uparrow}\oplus C^{\downarrow}\in\mathbb{C}(8)\oplus\mathbb{C}(8)$. These matrices are diagonal $8\times8$ block matrices due to the isomorphism $C\ell^+(8)\cong C\ell(7)\cong\mathbb{C}(8)\oplus\mathbb{C}(8)$. This means that the left and right adjoint actions of $C\ell(8)$ are a straightforward direct sum extension\footnote{Apart from the action of $\rho(J_i)$ which are not direct sum extensions.} of the $C\ell(6)$ left and right adjoint actions in \cite{Stoica2018},
\begin{eqnarray}
\mathbb{C}(8)\ni\rho(\Gamma_{\mu})&\mapsto& \rho(\Gamma_{\mu})\oplus\rho(\Gamma_{\mu})\in \mathbb{C}(8)\oplus\mathbb{C}(8),\\
\mathbb{C}(8)\ni\rho(T_i)&\mapsto& \rho(T_i)\oplus\rho(T_i)\in \mathbb{C}(8)\oplus\mathbb{C}(8),\\
\mathbb{C}(8)\ni\rho(J_i)&\mapsto &\rho(J_i)\oplus\rho(J_i)\in \mathbb{C}(8)\oplus\mathbb{C}(8).
\end{eqnarray}
For example, the weak generators function as follows on the first double spinors:
\begin{equation}
\rho(S_1^+)\oplus\rho(S_1^-)=\left(\begin{smallmatrix}
a_1&0&0&0&0&0&0&0 \\
a_2&0&0&0&0&0&0&0 \\
a_3&0&0&0&0&0&0&0 \\
a_4&0&0&0&0&0&0&0 \\
a_5&0&0&0&0&0&0&0 \\
a_6&0&0&0&0&0&0&0 \\
a_7&0&0&0&0&0&0&0 \\
a_8&0&0&0&0&0&0&0 \\
\end{smallmatrix}\right)\oplus\left(\begin{smallmatrix}
b_1&0&0&0&0&0&0&0 \\
b_2&0&0&0&0&0&0&0 \\
b_3&0&0&0&0&0&0&0 \\
b_4&0&0&0&0&0&0&0 \\
b_5&0&0&0&0&0&0&0 \\
b_6&0&0&0&0&0&0&0 \\
b_7&0&0&0&0&0&0&0 \\
b_8&0&0&0&0&0&0&0 \\
\end{smallmatrix}\right),
\end{equation}
where the $a$'s and $b$'s are complex numbers. We see that 
\begin{equation}
(\rho(T_1)\oplus\rho(T_1))(\rho(S_1^+)\oplus\rho(S_1^-))=\left(\begin{smallmatrix}
0&0&0&0&0&0&0&0 \\
0&0&0&0&0&0&0&0 \\
a_5&0&0&0&0&0&0&0 \\
a_6&0&0&0&0&0&0&0 \\
a_3&0&0&0&0&0&0&0 \\
a_4&0&0&0&0&0&0&0 \\
0&0&0&0&0&0&0&0 \\
0&0&0&0&0&0&0&0 \\
\end{smallmatrix}\right)\oplus\left(\begin{smallmatrix}
0&0&0&0&0&0&0&0 \\
0&0&0&0&0&0&0&0 \\
b_5&0&0&0&0&0&0&0 \\
b_6&0&0&0&0&0&0&0 \\
b_3&0&0&0&0&0&0&0 \\
b_4&0&0&0&0&0&0&0 \\
0&0&0&0&0&0&0&0 \\
0&0&0&0&0&0&0&0 \\
\end{smallmatrix}\right).
\end{equation}

Finally, we address the interaction between the actions of triality and the (right action) MTIS electrocolor symmetry group $U(3)_{\mathrm{ec}}$. The actions are independent. This is because triality stabilizes the exceptional Lie group $G_2$ pointwise. This means that for all $X\in G_2$ we have Trial$(X)=X$. Since $SU(3)_\mathrm{c}\subset G_2$ it follows that triality also stabilizes the color group pointwise. We write:
\begin{equation}
\{\rho(\Gamma_{\mu}),\rho(T_i),\rho(J_3)\}\triangleright \{A^{\uparrow}\oplus B^{\downarrow},C^{\uparrow}\oplus A^{\downarrow},B^{\uparrow}\oplus C^{\downarrow}\}\triangleleft\{\rho(\Lambda_1),\cdots,\rho(\Lambda_8),\rho(Q)\}.
\end{equation}
\subsection{Triality and intergenerational mixing}
\label{sec:ewmixing}
The action of triality on $\mathbb{C}(8)$ permutes through the different diagonal $\mathbb{C}(8)$ blocks, which physically corresponds to permuting through the three generations of fermions. Because each generation individually transforms as required (both via the left and right adjoint actions), a linear combination of all three generations will transform in an equivalent manner. In order to preserve unitarity, these linear combinations have to be part of a unitary matrix. If we therefore consider two suggestively labeled $3\times 3$ unitary matrices, $U_{\mathrm{PMNS}}$ and $U_{\mathrm{CKM}}$, we can then write:
\begin{eqnarray}
\left(\begin{array}{cccc}
\nu_{e,m} \\
\nu_{\mu,m} \\
\nu_{\tau,m}
\end{array}\right)=U_{\mathrm{PMNS}}\left(\begin{array}{cccc}
\nu_e \\
\nu_\mu \\
\nu_\tau
\end{array}\right),\quad\left(\begin{array}{cccc}
d_{s}^{(i)} \\
s_{s}^{(i)} \\
b_{s}^{(i)}
\end{array}\right)=U_{\mathrm{CKM}}\left(\begin{array}{cccc}
d^{(i)} \\
s^{(i)} \\
b^{(i)}
\end{array}\right),
\end{eqnarray}
where $\nu_{e,m}$, $d_s^{(i)}$ ($i=r,y,b$) e.t.c refers to the mixed states. However, in addition to $U_{\mathrm{PMNS}}$ and $U_{\mathrm{CKM}}$ giving neutrino and $(d,s,b)$ quark mixing, we can likewise write down additional mixing matrices for the remaining leptons and quarks. This would give more general flavour oscillations between $(e,\mu,\tau)$, $(u,c,t)$ or $(d,s,b)$. We have therefore shown that general mixing between flavours is possible. Additional insights/restrictions are required in order to obtain observed SM electroweak mixing.
\section{$C\ell(8)$ as the left adjoint algebra of the complex sedenions $\mathbb{C}\otimes\mathbb{S}$}
\label{sec:sed}
In a recent paper, we showed that three generations of leptons and quarks with unbroken gauge symmetry $SU(3)_c\times U(1)_{em}$ can be described using the algebra of complexified sedenions $\mathbb{C}\otimes\mathbb{S}$ \cite{gillard2019three}. One may wonder what the connection between that paper and the current paper is. This section addresses this question.

$C\ell(6)$ corresponds to the algebra of left (and right) adjoint actions of the complex octonions $\mathbb{C}\otimes\mathbb{O}$, and so we write
\begin{eqnarray}
(\mathbb{C}\otimes\mathbb{O})_L\cong C\ell(6).
\end{eqnarray}
The algebra of complex octonions $\mathbb{C}\otimes\mathbb{O}$ was first studied in relation to quark symmetries in \cite{gunaydin1973quark,gunaydin1974quark}. More recently, this algebra together with its adjoint algebras $C\ell(6)$, have also been studied in \cite{furey2016standard,dixon2010division,dixon1990derivation}. 

If, instead of $\mathbb{C}\otimes\mathbb{O}$ one considers the larger algebra $\mathbb{C}\otimes\mathbb{S}$, then the left (or right) adjoint algebra is likewise larger. In fact, now
\begin{equation}
(\mathbb{C}\otimes\mathbb{S})_L\cong C\ell(8),
\end{equation}
with $C\ell(8)$ being the algebra studied in the present paper. It was shown in  \cite{gillard2019three} that within $C\ell(8)$ one can represent three generations of fermions in terms of three $C\ell(6)$ subalgebras. These three subalgebras are not independent of one another but all share a common $C\ell(2)$ subalgebra. The general basis states of the minimal left ideals of $C\ell(6)$ are pinors, not spinors. However, pinors of $C\ell(n)$ can be regarded as spinors in $C\ell(n+1)\cong C\ell^+(n+2)$, and so the $C\ell(6)$ pinors correspond to physical states when they are considered in the larger algebra $C\ell(8)=(\mathbb{C}\otimes\mathbb{S})_L$, considered in the present paper.

A general element $A\in\mathbb{S}$ can be written in a canonical basis as
\begin{eqnarray}
A= \sum_{i=0}^{15} a_i  e_i= a_0+\sum_{i=1}^{15} a_i  e_i,\qquad a_i\in\mathbb{R},
\end{eqnarray}
where the basis elements satisfy the following multiplication rules
\begin{eqnarray}
\nonumber e_0&=&1,\qquad e_0e_i=e_ie_0=e_i,\\
e_1^2&=&e_2^2=...=e_{15}^2=-1,\\
\nonumber e_ie_j&=&-\delta_{ij}e_0+\gamma^k_{ij}e_k,
\end{eqnarray}
and the real structure constants $\gamma^k_{ij}$ are completely antisymmetric. Because $\mathbb{S}$ is a Cayley-Dickson algebra, one can also write $A\in\mathbb{S}$ in terms of a pair of complex octonion $O_1,O_2$, together with an anticommuting unit imaginary $e_8$. That is $A=O_1+e_8 O_2$. Then using the Witt basis $\{q_i, q_i^{\dagger}\}$ from Section \ref{sec:onegen}, we can write a minimal left $C\ell(6)$ ideal generated via the left adjoint actions of $O_1$, in terms of $\{q_i, q_i^{\dagger}\}$. A second minimal left $C\ell(6)$ ideal can be generated from the left adjoint actions of $e_8 O_2$ in terms of $\{r_i, r_i^{\dagger}\}$, where $r_i\equiv e_8q_i$. Both of these ideals contain a single generation of fermions of fixed (but opposite) spin. Each copy of $C\ell(6)$ can be written in terms of $\mathbb{C}(8)$, and so because
\begin{eqnarray}
\{q_i,r_j\}=\{q_i^{\dagger},r_j\}=\{q_i,r_j^{\dagger}\}=0,
\end{eqnarray}
together they can be expressed in terms of $\mathbb{C}(8)\oplus\mathbb{C}(8)$. We therefore have two physically distinct full sets of one generation of elementary fermion states which we could express in matrix representations in terms of $\mathbb{C}(8)\oplus\mathbb{C}(8)$. 

Furthermore, we observe that
\begin{equation}
\mathrm{Aut}(\mathbb{S})=\mathrm{Aut}(\mathbb{O})\times S_3
\end{equation}
and that $S_3$ is generated by triality \cite{lounesto2001caa}. Together with the isomorphism $(\mathbb{C}\otimes\mathbb{S})_L\cong C\ell(8)$ we find that the action of triality is as given in the previous section, and gives rise to three generations of fixed spin for the case of $C\ell(6)$ and three generations with spin degrees of freedom for the case of $C\ell(8)$.

For the $C\ell(8)$ case, only the subset of minimal left ideals contained in $C\ell^+(8)\cong\mathbb{C}(8)\oplus\mathbb{C}(8)\subset\mathbb{C}(16)$ correspond to physical quantum states. An analogous restriction is required when working with the sedenions. For arbitrary $a,b,c,d\in\mathbb{C}$ Consider
\begin{equation}
\left(\begin{smallmatrix}
a&b \\
c&d
\end{smallmatrix}\right)\left(\begin{smallmatrix}
O_1 \\
e_8O_2
\end{smallmatrix}\right)=\left(\begin{smallmatrix}
aO_1+be_8O_2 \\
cO_1+de_8O_2
\end{smallmatrix}\right).
\end{equation} 
The requirement of having only spinors states then means that one requires $b=c=0$. In this case the physical spinor space can be thought of as  $\mathbb{O}^2$, as is evident from the preceding equation.
\section{Concluding remarks}
\label{sec:Conclusions}
In this paper we have extended earlier work \cite{Stoica2018} where it was shown that for minimal left ideals in $\mathbb{C}(8)\cong C\ell(6)$ one gets one full generation of SM fermion states, albeit with no spin degrees of freedom. Accounting for the spin degrees of freedom, via the inclusion of an additional $C\ell(2)$ algebra, extends the relevant algebra from $C\ell(6) \cong \mathbb{C}(8)$ to $C\ell(8) \cong \mathbb{C}(16)$. The action of the $C\ell(2)_{\mathrm{isospin}}$ and $C\ell(2)_{\mathrm{spin}}$ factors on the minimal left ideals of $C\ell(8)$ is different. Whereas the isospin action preserves spinor spaces, the ladder spin operators do not.

The general basis states of the minimal left ideals of $C\ell(6)$ are pinors, not spinors. However, pinors of $C\ell(n)$ can be regarded as spinors in $C\ell(n+1)\cong C\ell^+(n+2)$. Unlike $C\ell(6)$, in $C\ell^+(8)\cong C\ell(7)$ the minimal left ideals correspond to actual $C\ell(8)$ algebraic spinors, $S_8$. The spinor spaces $S_8^\pm$ form the double spinor space $S_8^+\oplus S_8^-\equiv\mathbb{C}(8)\oplus\mathbb{C}(8)\cong C\ell(7)\cong C\ell^+(8)$. The $S_8^+$ double spinor subspace contains exclusively spin-up fermion states while the $S_8^-$ double spinor subspace contains exclusively spin-down fermion states.

The left adjoint action of this $C\ell(8)$ algebra accounts for the Dirac, weak isospin, and spin degrees of freedom and we write $C\ell(8)_L=C\ell(8)_{\mathrm{Dirac+isospin+spin}}$. However, only the spin-$z$ generator $J_z$ preserves the spinor spaces. The spin raising and lowering operators $J_\pm$ take spinors to pinors, and are thus unphysical. The physical states, corresponding to $C\ell(8)$ spinors, are preserved only by the subalgebra $C\ell(7)_{\mathrm{Dirac+isospin+spin-z}}\subset C\ell(8)_{\mathrm{Dirac+isospin+spin}}$. This explains why one observes physical $W^\pm$ bosons associated with weak isospin symmetry, but not analogous bosons for spin symmetry. The lack of an observed boson related to $J_z$ could be the result of there not being a $U(1)$ generator available to mix with.

The right adjoint action of the $C\ell(8)$ algebra gives the MTIS symmetry group $U(4)$. As with the left adjoint actions however, not all $U(4)$ transformations preserve the spinor space. The electrocolor subgroup $U(3)_{\mathrm{ec}}\subset U(4)$ of $U(4)$ is the maximal subgroup that does. This explains why there are only eight gluons in nature. 

The physically allowed parts of the left and right adjoint actions of $C\ell(8)$ therefore are:
\begin{equation}
C\ell(7)_{\mathrm{Dirac+isospin+spin-z}}\triangleright \{S_{1}^+\oplus S_{1}^-,\cdots,S_{8}^+\oplus S_{8}^-\}\triangleleft U(3)_{\mathrm{ec}},
\end{equation}
where $S_i^+$ and $S_i^-$ represent the $8\times 8$ spin up and spin down blocks respectively. These double spinors $S_{8,i}^+\oplus S_{8,i}^-$, $i=1,\cdots,8$, can be thought of as elements of $\mathbb{O}^2$, see the last chapter of \cite{lounesto2001caa}.

Unlike $C\ell(6)$, the larger algebra $C\ell(8)$ admits a triality automorphism. This automorphism of
$\mathrm{Spin}(8)$ permutes the two spinor and fundamental vector representations, all three of which are eight-dimensional. We have demonstrated that this triality automorphism extends all of the results from a single generation of fixed spin to three generations including spin degrees of freedom.
\section*{Appendix A}
\subsection*{Minimal left ideals of $C\ell(8)$}
\begin{equation}
\label{eqn:pinors1}
R_1=\left(\begin{smallmatrix}
p&pq_{32}&pq_{13}&pq_{21}&pq_{4321}&pq_{41}&pq_{42}&pq_{43} \\
q_{23}^\dagger p&q_{23}^\dagger pq_{32}&q_{23}^\dagger pq_{13}&q_{23}^\dagger pq_{21}&q_{23}^\dagger pq_{4321}&q_{23}^\dagger pq_{41}&q_{23}^\dagger pq_{42}&q_{23}^\dagger pq_{43} \\
q_{31}^\dagger p&q_{31}^\dagger pq_{32}&q_{31}^\dagger pq_{13}&q_{31}^\dagger pq_{21}&q_{31}^\dagger pq_{4321}&q_{31}^\dagger pq_{41}&q_{31}^\dagger pq_{42}&q_{31}^\dagger pq_{43} \\
q_{12}^\dagger p&q_{12}^\dagger pq_{32}&q_{12}^\dagger pq_{13}&q_{12}^\dagger pq_{21}&q_{12}^\dagger pq_{4321}&q_{12}^\dagger pq_{41}&q_{12}^\dagger pq_{42}&q_{12}^\dagger pq_{43} \\
q_{1234}^\dagger p&q_{1234}^\dagger pq_{32}&q_{1234}^\dagger pq_{13}&q_{1234}^\dagger pq_{21}&q_{1234}^\dagger pq_{4321}&q_{1234}^\dagger pq_{41}&q_{1234}^\dagger pq_{42}&q_{1234}^\dagger pq_{43} \\
q_{14}^\dagger p&q_{14}^\dagger pq_{32}&q_{14}^\dagger pq_{13}&q_{14}^\dagger pq_{21}&q_{14}^\dagger pq_{4321}&q_{14}^\dagger pq_{41}&q_{14}^\dagger pq_{42}&q_{14}^\dagger pq_{43} \\
q_{24}^\dagger p&q_{24}^\dagger pq_{32}&q_{24}^\dagger pq_{13}&q_{24}^\dagger pq_{21}&q_{24}^\dagger pq_{4321}&q_{24}^\dagger pq_{41}&q_{24}^\dagger pq_{42}&q_{24}^\dagger pq_{43} \\
q_{34}^\dagger p&q_{34}^\dagger pq_{32}&q_{34}^\dagger pq_{13}&q_{34}^\dagger pq_{21}&q_{34}^\dagger pq_{4321}&q_{34}^\dagger pq_{41}&q_{34}^\dagger pq_{42}&q_{34}^\dagger pq_{43} \\
q_{4}^\dagger p&q_{4}^\dagger pq_{32}&q_{4}^\dagger pq_{13}&q_{4}^\dagger pq_{21}&q_{4}^\dagger pq_{4321}&q_{4}^\dagger pq_{41}&q_{4}^\dagger pq_{42}&q_{4}^\dagger pq_{43} \\
q_{234}^\dagger p&q_{234}^\dagger pq_{32}&q_{234}^\dagger pq_{13}&q_{234}^\dagger pq_{21}&q_{234}^\dagger pq_{4321}&q_{234}^\dagger pq_{41}&q_{234}^\dagger pq_{42}&q_{234}^\dagger pq_{43} \\
q_{314}^\dagger p&q_{314}^\dagger pq_{32}&q_{314}^\dagger pq_{13}&q_{314}^\dagger pq_{21}&q_{314}^\dagger pq_{4321}&q_{314}^\dagger pq_{41}&q_{314}^\dagger pq_{42}&q_{314}^\dagger pq_{43} \\
q_{124}^\dagger p&q_{124}^\dagger pq_{32}&q_{124}^\dagger pq_{13}&q_{124}^\dagger pq_{21}&q_{124}^\dagger pq_{4321}&q_{124}^\dagger pq_{41}&q_{124}^\dagger pq_{42}&q_{124}^\dagger pq_{43} \\
q_{123}^\dagger p&q_{123}^\dagger pq_{32}&q_{123}^\dagger pq_{13}&q_{123}^\dagger pq_{21}&q_{123}^\dagger pq_{4321}&q_{123}^\dagger pq_{41}&q_{123}^\dagger pq_{42}&q_{123}^\dagger pq_{43} \\
q_{1}^\dagger p&q_{1}^\dagger pq_{32}&q_{1}^\dagger pq_{13}&q_{1}^\dagger pq_{21}&q_{1}^\dagger pq_{4321}&q_{1}^\dagger pq_{41}&q_{1}^\dagger pq_{42}&q_{1}^\dagger pq_{43} \\
q_{2}^\dagger p&q_{2}^\dagger pq_{32}&q_{2}^\dagger pq_{13}&q_{2}^\dagger pq_{21}&q_{2}^\dagger pq_{4321}&q_{2}^\dagger pq_{41}&q_{2}^\dagger pq_{42}&q_{2}^\dagger pq_{43} \\
q_{3}^\dagger p&q_{3}^\dagger pq_{32}&q_{3}^\dagger pq_{13}&q_{3}^\dagger pq_{21}&q_{3}^\dagger pq_{4321}&q_{3}^\dagger pq_{41}&q_{3}^\dagger pq_{42}&q_{3}^\dagger pq_{43} 
\end{smallmatrix}\right),
\end{equation}
\begin{equation}
\label{eqn:pinors2}
R_2=\left(\begin{smallmatrix}
pq_{4}&pq_{432}&pq_{413}&pq_{421}&pq_{321}&pq_{1}&pq_{2}&pq_{3} \\
q_{23}^\dagger pq_4&q_{23}^\dagger pq_{432}&q_{23}^\dagger pq_{413}&q_{23}^\dagger pq_{421}&q_{23}^\dagger pq_{321}&q_{23}^\dagger pq_{1}&q_{23}^\dagger pq_{2}&q_{23}^\dagger pq_{3} \\
q_{31}^\dagger pq_{4}&q_{31}^\dagger pq_{432}&q_{31}^\dagger pq_{413}&q_{31}^\dagger pq_{421}&q_{31}^\dagger pq_{321}&q_{31}^\dagger pq_{1}&q_{31}^\dagger pq_{2}&q_{31}^\dagger pq_{3} \\
q_{12}^\dagger pq_{4}&q_{12}^\dagger pq_{432}&q_{12}^\dagger pq_{413}&q_{12}^\dagger pq_{421}&q_{12}^\dagger pq_{321}&q_{12}^\dagger pq_{1}&q_{12}^\dagger pq_{2}&q_{12}^\dagger pq_{3} \\
q_{1234}^\dagger pq_{4}&q_{1234}^\dagger pq_{432}&q_{1234}^\dagger pq_{413}&q_{1234}^\dagger pq_{421}&q_{1234}^\dagger pq_{321}&q_{1234}^\dagger pq_{1}&q_{1234}^\dagger pq_{2}&q_{1234}^\dagger pq_{3} \\
q_{14}^\dagger pq_{4}&q_{14}^\dagger pq_{432}&q_{14}^\dagger pq_{413}&q_{14}^\dagger pq_{421}&q_{14}^\dagger pq_{321}&q_{14}^\dagger pq_{1}&q_{14}^\dagger pq_{2}&q_{14}^\dagger pq_{3} \\
q_{24}^\dagger pq_{4}&q_{24}^\dagger pq_{432}&q_{24}^\dagger pq_{413}&q_{24}^\dagger pq_{421}&q_{24}^\dagger pq_{321}&q_{24}^\dagger pq_{1}&q_{24}^\dagger pq_{2}&q_{24}^\dagger pq_{3} \\
q_{34}^\dagger pq_{4}&q_{34}^\dagger pq_{432}&q_{34}^\dagger pq_{413}&q_{34}^\dagger pq_{421}&q_{34}^\dagger pq_{321}&q_{34}^\dagger pq_{1}&q_{34}^\dagger pq_{2}&q_{34}^\dagger pq_{3} \\
q_{4}^\dagger pq_{4}&q_{4}^\dagger pq_{432}&q_{4}^\dagger pq_{413}&q_{4}^\dagger pq_{421}&q_{4}^\dagger pq_{321}&q_{4}^\dagger pq_{1}&q_{4}^\dagger pq_{2}&q_{4}^\dagger pq_{3} \\
q_{234}^\dagger pq_{4}&q_{234}^\dagger pq_{432}&q_{234}^\dagger pq_{413}&q_{234}^\dagger pq_{421}&q_{234}^\dagger pq_{321}&q_{234}^\dagger pq_{1}&q_{234}^\dagger pq_{2}&q_{234}^\dagger pq_{3} \\
q_{314}^\dagger pq_{4}&q_{314}^\dagger pq_{432}&q_{314}^\dagger pq_{413}&q_{314}^\dagger pq_{421}&q_{314}^\dagger pq_{321}&q_{314}^\dagger pq_{1}&q_{314}^\dagger pq_{2}&q_{314}^\dagger pq_{3} \\
q_{124}^\dagger pq_{4}&q_{124}^\dagger pq_{432}&q_{124}^\dagger pq_{413}&q_{124}^\dagger pq_{421}&q_{124}^\dagger pq_{321}&q_{124}^\dagger pq_{1}&q_{124}^\dagger pq_{2}&q_{124}^\dagger pq_{3} \\
q_{123}^\dagger pq_{4}&q_{123}^\dagger pq_{432}&q_{123}^\dagger pq_{413}&q_{123}^\dagger pq_{421}&q_{123}^\dagger pq_{321}&q_{123}^\dagger pq_{1}&q_{123}^\dagger pq_{2}&q_{123}^\dagger pq_{3} \\
q_{1}^\dagger pq_{4}&q_{1}^\dagger pq_{432}&q_{1}^\dagger pq_{413}&q_{1}^\dagger pq_{421}&q_{1}^\dagger pq_{321}&q_{1}^\dagger pq_{1}&q_{1}^\dagger pq_{2}&q_{1}^\dagger pq_{3} \\
q_{2}^\dagger pq_{4}&q_{2}^\dagger pq_{432}&q_{2}^\dagger pq_{413}&q_{2}^\dagger pq_{421}&q_{2}^\dagger pq_{321}&q_{2}^\dagger pq_{1}&q_{2}^\dagger pq_{2}&q_{2}^\dagger pq_{3} \\
q_{3}^\dagger pq_{4}&q_{3}^\dagger pq_{432}&q_{3}^\dagger pq_{413}&q_{3}^\dagger pq_{421}&q_{3}^\dagger pq_{321}&q_{3}^\dagger pq_{1}&q_{3}^\dagger pq_{2}&q_{3}^\dagger pq_{3} 
\end{smallmatrix}\right),
\end{equation}
\subsection*{Spinors of $C\ell(8)$}
\begin{equation}
\label{eqn:spinors1}
R_{1,s}=\left(\begin{smallmatrix}
p&pq_{32}&pq_{13}&pq_{21}&pq_{4321}&pq_{41}&pq_{42}&pq_{43} \\
q_{23}^\dagger p&q_{23}^\dagger pq_{32}&q_{23}^\dagger pq_{13}&q_{23}^\dagger pq_{21}&q_{23}^\dagger pq_{4321}&q_{23}^\dagger pq_{41}&q_{23}^\dagger pq_{42}&q_{23}^\dagger pq_{43} \\
q_{31}^\dagger p&q_{31}^\dagger pq_{32}&q_{31}^\dagger pq_{13}&q_{31}^\dagger pq_{21}&q_{31}^\dagger pq_{4321}&q_{31}^\dagger pq_{41}&q_{31}^\dagger pq_{42}&q_{31}^\dagger pq_{43} \\
q_{12}^\dagger p&q_{12}^\dagger pq_{32}&q_{12}^\dagger pq_{13}&q_{12}^\dagger pq_{21}&q_{12}^\dagger pq_{4321}&q_{12}^\dagger pq_{41}&q_{12}^\dagger pq_{42}&q_{12}^\dagger pq_{43} \\
q_{1234}^\dagger p&q_{1234}^\dagger pq_{32}&q_{1234}^\dagger pq_{13}&q_{1234}^\dagger pq_{21}&q_{1234}^\dagger pq_{4321}&q_{1234}^\dagger pq_{41}&q_{1234}^\dagger pq_{42}&q_{1234}^\dagger pq_{43} \\
q_{14}^\dagger p&q_{14}^\dagger pq_{32}&q_{14}^\dagger pq_{13}&q_{14}^\dagger pq_{21}&q_{14}^\dagger pq_{4321}&q_{14}^\dagger pq_{41}&q_{14}^\dagger pq_{42}&q_{14}^\dagger pq_{43} \\
q_{24}^\dagger p&q_{24}^\dagger pq_{32}&q_{24}^\dagger pq_{13}&q_{24}^\dagger pq_{21}&q_{24}^\dagger pq_{4321}&q_{24}^\dagger pq_{41}&q_{24}^\dagger pq_{42}&q_{24}^\dagger pq_{43} \\
q_{34}^\dagger p&q_{34}^\dagger pq_{32}&q_{34}^\dagger pq_{13}&q_{34}^\dagger pq_{21}&q_{34}^\dagger pq_{4321}&q_{34}^\dagger pq_{41}&q_{34}^\dagger pq_{42}&q_{34}^\dagger pq_{43} \\
0&0&0&0&0&0&0&0 \\
0&0&0&0&0&0&0&0 \\
0&0&0&0&0&0&0&0 \\
0&0&0&0&0&0&0&0 \\
0&0&0&0&0&0&0&0 \\
0&0&0&0&0&0&0&0 \\
0&0&0&0&0&0&0&0 \\
0&0&0&0&0&0&0&0 
\end{smallmatrix}\right),
\end{equation}
\begin{equation}
\label{eqn:pinors2}
R_{2,s}=\left(\begin{smallmatrix}
0&0&0&0&0&0&0&0 \\
0&0&0&0&0&0&0&0 \\
0&0&0&0&0&0&0&0 \\
0&0&0&0&0&0&0&0 \\
0&0&0&0&0&0&0&0 \\
0&0&0&0&0&0&0&0 \\
0&0&0&0&0&0&0&0 \\
0&0&0&0&0&0&0&0 \\
q_{4}^\dagger pq_{4}&q_{4}^\dagger pq_{432}&q_{4}^\dagger pq_{413}&q_{4}^\dagger pq_{421}&q_{4}^\dagger pq_{321}&q_{4}^\dagger pq_{1}&q_{4}^\dagger pq_{2}&q_{4}^\dagger pq_{3} \\
q_{234}^\dagger pq_{4}&q_{234}^\dagger pq_{432}&q_{234}^\dagger pq_{413}&q_{234}^\dagger pq_{421}&q_{234}^\dagger pq_{321}&q_{234}^\dagger pq_{1}&q_{234}^\dagger pq_{2}&q_{234}^\dagger pq_{3} \\
q_{314}^\dagger pq_{4}&q_{314}^\dagger pq_{432}&q_{314}^\dagger pq_{413}&q_{314}^\dagger pq_{421}&q_{314}^\dagger pq_{321}&q_{314}^\dagger pq_{1}&q_{314}^\dagger pq_{2}&q_{314}^\dagger pq_{3} \\
q_{124}^\dagger pq_{4}&q_{124}^\dagger pq_{432}&q_{124}^\dagger pq_{413}&q_{124}^\dagger pq_{421}&q_{124}^\dagger pq_{321}&q_{124}^\dagger pq_{1}&q_{124}^\dagger pq_{2}&q_{124}^\dagger pq_{3} \\
q_{123}^\dagger pq_{4}&q_{123}^\dagger pq_{432}&q_{123}^\dagger pq_{413}&q_{123}^\dagger pq_{421}&q_{123}^\dagger pq_{321}&q_{123}^\dagger pq_{1}&q_{123}^\dagger pq_{2}&q_{123}^\dagger pq_{3} \\
q_{1}^\dagger pq_{4}&q_{1}^\dagger pq_{432}&q_{1}^\dagger pq_{413}&q_{1}^\dagger pq_{421}&q_{1}^\dagger pq_{321}&q_{1}^\dagger pq_{1}&q_{1}^\dagger pq_{2}&q_{1}^\dagger pq_{3} \\
q_{2}^\dagger pq_{4}&q_{2}^\dagger pq_{432}&q_{2}^\dagger pq_{413}&q_{2}^\dagger pq_{421}&q_{2}^\dagger pq_{321}&q_{2}^\dagger pq_{1}&q_{2}^\dagger pq_{2}&q_{2}^\dagger pq_{3} \\
q_{3}^\dagger pq_{4}&q_{3}^\dagger pq_{432}&q_{3}^\dagger pq_{413}&q_{3}^\dagger pq_{421}&q_{3}^\dagger pq_{321}&q_{3}^\dagger pq_{1}&q_{3}^\dagger pq_{2}&q_{3}^\dagger pq_{3} 
\end{smallmatrix}\right),
\end{equation}

\subsection*{Full pinor spaces}
\begin{equation}
\label{eqn:three0}
\mathbb{P}_1\equiv\left(\begin{smallmatrix}
\nu_{e,R_1}^\uparrow&u_{R_1}^{r\uparrow}&u_{R_1}^{y\uparrow}&u_{R_1}^{b\uparrow}&e^{+\uparrow}_{L1}&\bar{d}_{L1}^{\bar{r}\uparrow}&\bar{d}_{L1}^{\bar{y}\uparrow}&\bar{d}_{L1}^{\bar{b}\uparrow}&\nu_{\tau,R_1}^\uparrow&t_{R_1}^{r\uparrow}&t_{R_1}^{y\uparrow}&t_{R_1}^{b\uparrow}&\tau^{+\uparrow}_{L1}&\bar{b}_{L1}^{\bar{r}\uparrow}&\bar{b}_{L1}^{\bar{y}\uparrow}&\bar{b}_{L1}^{\bar{b}\uparrow} \\
\nu_{e,R_2}^\uparrow&u_{R_2}^{r\uparrow}&u_{R_2}^{y\uparrow}&u_{R_2}^{b\uparrow}&e^{+\uparrow}_{L2}&\bar{d}_{L2}^{\bar{r}\uparrow}&\bar{d}_{L2}^{\bar{y}\uparrow}&\bar{d}_{L2}^{\bar{b}\uparrow}&\nu_{\tau,R_2}^\uparrow&t_{R_2}^{r\uparrow}&t_{R_2}^{y\uparrow}&t_{R_2}^{b\uparrow}&\tau^{+\uparrow}_{L2}&\bar{b}_{L2}^{\bar{r}\uparrow}&\bar{b}_{L2}^{\bar{y}\uparrow}&\bar{b}_{L2}^{\bar{b}\uparrow} \\
\nu_{e,L_1}^\uparrow&u_{R_1}^{r\uparrow}&u_{L_1}^{y\uparrow}&u_{L_1}^{b\uparrow}&e^{+\uparrow}_{R1}&\bar{d}_{R1}^{\bar{r}\uparrow}&\bar{d}_{R1}^{\bar{y}\uparrow}&\bar{d}_{R1}^{\bar{b}\uparrow}&\nu_{\tau,L_1}^\uparrow&t_{R_1}^{r\uparrow}&t_{L_1}^{y\uparrow}&t_{L_1}^{b\uparrow}&\tau^{+\uparrow}_{R1}&\bar{b}_{R1}^{\bar{r}\uparrow}&\bar{b}_{R1}^{\bar{y}\uparrow}&\bar{b}_{R1}^{\bar{b}\uparrow} \\
\nu_{e,L_2}^\uparrow&u_{R_2}^{r\uparrow}&u_{L_2}^{y\uparrow}&u_{L_2}^{b\uparrow}&e^{+\uparrow}_{R2}&\bar{d}_{R2}^{\bar{r}\uparrow}&\bar{d}_{R2}^{\bar{y}\uparrow}&\bar{d}_{R2}^{\bar{b}\uparrow}&\nu_{\tau,L_2}^\uparrow&t_{R_2}^{r\uparrow}&t_{L_2}^{y\uparrow}&t_{L_2}^{b\uparrow}&\tau^{+\uparrow}_{R2}&\bar{b}_{R2}^{\bar{r}\uparrow}&\bar{b}_{R2}^{\bar{y}\uparrow}&\bar{b}_{R2}^{\bar{b}\uparrow}  \\
e^{-\uparrow}_{L1}&d_{L1}^{r\uparrow}&d_{L1}^{y\uparrow}&d_{L1}^{b\uparrow}&\bar{\nu}_{e,R1}^\uparrow&\bar{u}_{R1}^{\bar{r}\uparrow}&\bar{u}_{R1}^{\bar{y}\uparrow}&\bar{u}_{R1}^{\bar{b}\uparrow}&\tau^{-\uparrow}_{L1}&b_{L1}^{r\uparrow}&b_{L1}^{y\uparrow}&b_{L1}^{b\uparrow}&\bar{\nu}_{\tau,R1}^\uparrow&\bar{t}_{R1}^{\bar{r}\uparrow}&\bar{t}_{R1}^{\bar{y}\uparrow}&\bar{t}_{R1}^{\bar{b}\uparrow} \\
e^{-\uparrow}_{L2}&d_{L2}^{r\uparrow}&d_{L2}^{y\uparrow}&d_{L2}^{b\uparrow}&\bar{\nu}_{e,R2}^\uparrow&\bar{u}_{R2}^{\bar{r}\uparrow}&\bar{u}_{R2}^{\bar{y}\uparrow}&\bar{u}_{R2}^{\bar{b}\uparrow}&\tau^{-\uparrow}_{L2}&b_{L2}^{r\uparrow}&b_{L2}^{y\uparrow}&b_{L2}^{b\uparrow}&\bar{\nu}_{\tau,R2}^\uparrow&\bar{t}_{R2}^{\bar{r}\uparrow}&\bar{t}_{R2}^{\bar{y}\uparrow}&\bar{t}_{R2}^{\bar{b}\uparrow} \\
e^{-\uparrow}_{R1}&d_{R1}^{r\uparrow}&d_{R1}^{y\uparrow}&d_{R1}^{b\uparrow}&\bar{\nu}_{e,L1}^\uparrow&\bar{u}_{L1}^{\bar{r}\uparrow}&\bar{u}_{L1}^{\bar{y}\uparrow}&\bar{u}_{L1}^{\bar{b}\uparrow}&\tau^{-\uparrow}_{R1}&b_{R1}^{r\uparrow}&b_{R1}^{y\uparrow}&b_{R1}^{b\uparrow}&\bar{\nu}_{\tau,L1}^\uparrow&\bar{t}_{L1}^{\bar{r}\uparrow}&\bar{t}_{L1}^{\bar{y}\uparrow}&\bar{t}_{L1}^{\bar{b}\uparrow} \\
e^{-\uparrow}_{R2}&d_{R2}^{r\uparrow}&d_{R2}^{y\uparrow}&d_{R2}^{b\uparrow}&\bar{\nu}_{e,L2}^\uparrow&\bar{u}_{L2}^{\bar{r}\uparrow}&\bar{u}_{L2}^{\bar{y}\uparrow}&\bar{u}_{L2}^{\bar{b}\uparrow}&\tau^{-\uparrow}_{R2}&b_{R2}^{r\uparrow}&b_{R2}^{y\uparrow}&b_{R2}^{b\uparrow}&\bar{\nu}_{\tau,L2}^\uparrow&\bar{t}_{L2}^{\bar{r}\uparrow}&\bar{t}_{L2}^{\bar{y}\uparrow}&\bar{t}_{L2}^{\bar{b}\uparrow} \\
\nu_{e,R_1}^\downarrow&u_{R_1}^{r\downarrow}&u_{R_1}^{y\downarrow}&u_{R_1}^{b\downarrow}&e^{+\downarrow}_{L1}&\bar{d}_{L1}^{\bar{r}\downarrow}&\bar{d}_{L1}^{\bar{y}\downarrow}&\bar{d}_{L1}^{\bar{b}\downarrow}&\nu_{\tau,R_1}^\downarrow&t_{R_1}^{r\downarrow}&t_{R_1}^{y\downarrow}&t_{R_1}^{b\downarrow}&\tau^{+\downarrow}_{L1}&\bar{b}_{L1}^{\bar{r}\downarrow}&\bar{b}_{L1}^{\bar{y}\downarrow}&\bar{b}_{L1}^{\bar{b}\downarrow} \\
\nu_{e,R_2}^\downarrow&u_{R_2}^{r\downarrow}&u_{R_2}^{y\downarrow}&u_{R_2}^{b\downarrow}&e^{+\downarrow}_{L2}&\bar{d}_{L2}^{\bar{r}\downarrow}&\bar{d}_{L2}^{\bar{y}\downarrow}&\bar{d}_{L2}^{\bar{b}\downarrow}&\nu_{\tau,R_2}^\downarrow&t_{R_2}^{r\downarrow}&t_{R_2}^{y\downarrow}&t_{R_2}^{b\downarrow}&\tau^{+\downarrow}_{L2}&\bar{b}_{L2}^{\bar{r}\downarrow}&\bar{b}_{L2}^{\bar{y}\downarrow}&\bar{b}_{L2}^{\bar{b}\downarrow} \\
\nu_{e,L_1}^\downarrow&u_{R_1}^{r\downarrow}&u_{L_1}^{y\downarrow}&u_{L_1}^{b\downarrow}&e^{+\downarrow}_{R1}&\bar{d}_{R1}^{\bar{r}\downarrow}&\bar{d}_{R1}^{\bar{y}\downarrow}&\bar{d}_{R1}^{\bar{b}\downarrow}&\nu_{\tau,L_1}^\downarrow&t_{R_1}^{r\downarrow}&t_{L_1}^{y\downarrow}&t_{L_1}^{b\downarrow}&\tau^{+\downarrow}_{R1}&\bar{b}_{R1}^{\bar{r}\downarrow}&\bar{b}_{R1}^{\bar{y}\downarrow}&\bar{b}_{R1}^{\bar{b}\downarrow} \\
\nu_{e,L_2}^\downarrow&u_{R_2}^{r\downarrow}&u_{L_2}^{y\downarrow}&u_{L_2}^{b\downarrow}&e^{+\downarrow}_{R2}&\bar{d}_{R2}^{\bar{r}\downarrow}&\bar{d}_{R2}^{\bar{y}\downarrow}&\bar{d}_{R2}^{\bar{b}\downarrow}&\nu_{\tau,L_2}^\downarrow&t_{R_2}^{r\downarrow}&t_{L_2}^{y\downarrow}&t_{L_2}^{b\downarrow}&\tau^{+\downarrow}_{R2}&\bar{b}_{R2}^{\bar{r}\downarrow}&\bar{b}_{R2}^{\bar{y}\downarrow}&\bar{b}_{R2}^{\bar{b}\downarrow}  \\
e^{-\downarrow}_{L1}&d_{L1}^{r\downarrow}&d_{L1}^{y\downarrow}&d_{L1}^{b\downarrow}&\bar{\nu}_{e,R1}^\downarrow&\bar{u}_{R1}^{\bar{r}\downarrow}&\bar{u}_{R1}^{\bar{y}\downarrow}&\bar{u}_{R1}^{\bar{b}\downarrow}&\tau^{-\downarrow}_{L1}&b_{L1}^{r\downarrow}&b_{L1}^{y\downarrow}&b_{L1}^{b\downarrow}&\bar{\nu}_{\tau,R1}^\downarrow&\bar{t}_{R1}^{\bar{r}\downarrow}&\bar{t}_{R1}^{\bar{y}\downarrow}&\bar{t}_{R1}^{\bar{b}\downarrow} \\
e^{-\downarrow}_{L2}&d_{L2}^{r\downarrow}&d_{L2}^{y\downarrow}&d_{L2}^{b\downarrow}&\bar{\nu}_{e,R2}^\downarrow&\bar{u}_{R2}^{\bar{r}\downarrow}&\bar{u}_{R2}^{\bar{y}\downarrow}&\bar{u}_{R2}^{\bar{b}\downarrow}&\tau^{-\downarrow}_{L2}&b_{L2}^{r\downarrow}&b_{L2}^{y\downarrow}&b_{L2}^{b\downarrow}&\bar{\nu}_{\tau,R2}^\downarrow&\bar{t}_{R2}^{\bar{r}\downarrow}&\bar{t}_{R2}^{\bar{y}\downarrow}&\bar{t}_{R2}^{\bar{b}\downarrow} \\
e^{-\downarrow}_{R1}&d_{R1}^{r\downarrow}&d_{R1}^{y\downarrow}&d_{R1}^{b\downarrow}&\bar{\nu}_{e,L1}^\downarrow&\bar{u}_{L1}^{\bar{r}\downarrow}&\bar{u}_{L1}^{\bar{y}\downarrow}&\bar{u}_{L1}^{\bar{b}\downarrow}&\tau^{-\downarrow}_{R1}&b_{R1}^{r\downarrow}&b_{R1}^{y\downarrow}&b_{R1}^{b\downarrow}&\bar{\nu}_{\tau,L1}^\downarrow&\bar{t}_{L1}^{\bar{r}\downarrow}&\bar{t}_{L1}^{\bar{y}\downarrow}&\bar{t}_{L1}^{\bar{b}\downarrow} \\
e^{-\downarrow}_{R2}&d_{R2}^{r\downarrow}&d_{R2}^{y\downarrow}&d_{R2}^{b\downarrow}&\bar{\nu}_{e,L2}^\downarrow&\bar{u}_{L2}^{\bar{r}\downarrow}&\bar{u}_{L2}^{\bar{y}\downarrow}&\bar{u}_{L2}^{\bar{b}\downarrow}&\tau^{-\downarrow}_{R2}&b_{R2}^{r\downarrow}&b_{R2}^{y\downarrow}&b_{R2}^{b\downarrow}&\bar{\nu}_{\tau,L2}^\downarrow&\bar{t}_{L2}^{\bar{r}\downarrow}&\bar{t}_{L2}^{\bar{y}\downarrow}&\bar{t}_{L2}^{\bar{b}\downarrow} \\
\end{smallmatrix}\right),
\end{equation}
\begin{equation}
\label{eqn:three0}
\mathbb{P}_2\equiv\left(\begin{smallmatrix}
\nu_{\mu,R_1}^\uparrow&c_{R_1}^{r\uparrow}&c_{R_1}^{y\uparrow}&c_{R_1}^{b\uparrow}&\mu^{+\uparrow}_{L1}&\bar{s}_{L1}^{\bar{r}\uparrow}&\bar{s}_{L1}^{\bar{y}\uparrow}&\bar{s}_{L1}^{\bar{b}\uparrow}&\nu_{e,R_1}^\uparrow&u_{R_1}^{r\uparrow}&u_{R_1}^{y\uparrow}&u_{R_1}^{b\uparrow}&e^{+\uparrow}_{L1}&\bar{d}_{L1}^{\bar{r}\uparrow}&\bar{d}_{L1}^{\bar{y}\uparrow}&\bar{d}_{L1}^{\bar{b}\uparrow} \\
\nu_{\mu,R_2}^\uparrow&c_{R_2}^{r\uparrow}&c_{R_2}^{y\uparrow}&c_{R_2}^{b\uparrow}&\mu^{+\uparrow}_{L2}&\bar{s}_{L2}^{\bar{r}\uparrow}&\bar{s}_{L2}^{\bar{y}\uparrow}&\bar{s}_{L2}^{\bar{b}\uparrow}&\nu_{e,R_2}^\uparrow&u_{R_2}^{r\uparrow}&u_{R_2}^{y\uparrow}&u_{R_2}^{b\uparrow}&e^{+\uparrow}_{L2}&\bar{d}_{L2}^{\bar{r}\uparrow}&\bar{d}_{L2}^{\bar{y}\uparrow}&\bar{d}_{L2}^{\bar{b}\uparrow} \\
\nu_{\mu,L_1}^\uparrow&c_{R_1}^{r\uparrow}&c_{L_1}^{y\uparrow}&c_{L_1}^{b\uparrow}&\mu^{+\uparrow}_{R1}&\bar{s}_{R1}^{\bar{r}\uparrow}&\bar{s}_{R1}^{\bar{y}\uparrow}&\bar{s}_{R1}^{\bar{b}\uparrow}&\nu_{e,L_1}^\uparrow&u_{R_1}^{r\uparrow}&u_{L_1}^{y\uparrow}&u_{L_1}^{b\uparrow}&e^{+\uparrow}_{R1}&\bar{d}_{R1}^{\bar{r}\uparrow}&\bar{d}_{R1}^{\bar{y}\uparrow}&\bar{d}_{R1}^{\bar{b}\uparrow} \\
\nu_{\mu,L_2}^\uparrow&c_{R_2}^{r\uparrow}&c_{L_2}^{y\uparrow}&c_{L_2}^{b\uparrow}&\mu^{+\uparrow}_{R2}&\bar{s}_{R2}^{\bar{r}\uparrow}&\bar{s}_{R2}^{\bar{y}\uparrow}&\bar{s}_{R2}^{\bar{b}\uparrow}&\nu_{e,L_2}^\uparrow&u_{R_2}^{r\uparrow}&u_{L_2}^{y\uparrow}&u_{L_2}^{b\uparrow}&e^{+\uparrow}_{R2}&\bar{d}_{R2}^{\bar{r}\uparrow}&\bar{d}_{R2}^{\bar{y}\uparrow}&\bar{d}_{R2}^{\bar{b}\uparrow}  \\
\mu^{-\uparrow}_{L1}&s_{L1}^{r\uparrow}&s_{L1}^{y\uparrow}&s_{L1}^{b\uparrow}&\bar{\nu}_{\mu,R1}^\uparrow&\bar{c}_{R1}^{\bar{r}\uparrow}&\bar{c}_{R1}^{\bar{y}\uparrow}&\bar{c}_{R1}^{\bar{b}\uparrow}&e^{-\uparrow}_{L1}&d_{L1}^{r\uparrow}&d_{L1}^{y\uparrow}&d_{L1}^{b\uparrow}&\bar{\nu}_{e,R1}^\uparrow&\bar{u}_{R1}^{\bar{r}\uparrow}&\bar{u}_{R1}^{\bar{y}\uparrow}&\bar{u}_{R1}^{\bar{b}\uparrow} \\
\mu^{-\uparrow}_{L2}&s_{L2}^{r\uparrow}&s_{L2}^{y\uparrow}&s_{L2}^{b\uparrow}&\bar{\nu}_{\mu,R2}^\uparrow&\bar{c}_{R2}^{\bar{r}\uparrow}&\bar{c}_{R2}^{\bar{y}\uparrow}&\bar{c}_{R2}^{\bar{b}\uparrow}&e^{-\uparrow}_{L2}&d_{L2}^{r\uparrow}&d_{L2}^{y\uparrow}&d_{L2}^{b\uparrow}&\bar{\nu}_{e,R2}^\uparrow&\bar{u}_{R2}^{\bar{r}\uparrow}&\bar{u}_{R2}^{\bar{y}\uparrow}&\bar{u}_{R2}^{\bar{b}\uparrow} \\
\mu^{-\uparrow}_{R1}&s_{R1}^{r\uparrow}&s_{R1}^{y\uparrow}&s_{R1}^{b\uparrow}&\bar{\nu}_{\mu,L1}^\uparrow&\bar{c}_{L1}^{\bar{r}\uparrow}&\bar{c}_{L1}^{\bar{y}\uparrow}&\bar{c}_{L1}^{\bar{b}\uparrow}&e^{-\uparrow}_{R1}&d_{R1}^{r\uparrow}&d_{R1}^{y\uparrow}&d_{R1}^{b\uparrow}&\bar{\nu}_{e,L1}^\uparrow&\bar{u}_{L1}^{\bar{r}\uparrow}&\bar{u}_{L1}^{\bar{y}\uparrow}&\bar{u}_{L1}^{\bar{b}\uparrow} \\
\mu^{-\uparrow}_{R2}&s_{R2}^{r\uparrow}&s_{R2}^{y\uparrow}&s_{R2}^{b\uparrow}&\bar{\nu}_{\mu,L2}^\uparrow&\bar{c}_{L2}^{\bar{r}\uparrow}&\bar{c}_{L2}^{\bar{y}\uparrow}&\bar{c}_{L2}^{\bar{b}\uparrow}&e^{-\uparrow}_{R2}&d_{R2}^{r\uparrow}&d_{R2}^{y\uparrow}&d_{R2}^{b\uparrow}&\bar{\nu}_{e,L2}^\uparrow&\bar{u}_{L2}^{\bar{r}\uparrow}&\bar{u}_{L2}^{\bar{y}\uparrow}&\bar{u}_{L2}^{\bar{b}\uparrow} \\
\nu_{\mu,R_1}^\downarrow&c_{R_1}^{r\downarrow}&c_{R_1}^{y\downarrow}&c_{R_1}^{b\downarrow}&\mu^{+\downarrow}_{L1}&\bar{s}_{L1}^{\bar{r}\downarrow}&\bar{s}_{L1}^{\bar{y}\downarrow}&\bar{s}_{L1}^{\bar{b}\downarrow}&\nu_{e,R_1}^\downarrow&u_{R_1}^{r\downarrow}&u_{R_1}^{y\downarrow}&u_{R_1}^{b\downarrow}&e^{+\downarrow}_{L1}&\bar{d}_{L1}^{\bar{r}\downarrow}&\bar{d}_{L1}^{\bar{y}\downarrow}&\bar{d}_{L1}^{\bar{b}\downarrow} \\
\nu_{\mu,R_2}^\downarrow&c_{R_2}^{r\downarrow}&c_{R_2}^{y\downarrow}&c_{R_2}^{b\downarrow}&\mu^{+\downarrow}_{L2}&\bar{s}_{L2}^{\bar{r}\downarrow}&\bar{s}_{L2}^{\bar{y}\downarrow}&\bar{s}_{L2}^{\bar{b}\downarrow}&\nu_{e,R_2}^\downarrow&u_{R_2}^{r\downarrow}&u_{R_2}^{y\downarrow}&u_{R_2}^{b\downarrow}&e^{+\downarrow}_{L2}&\bar{d}_{L2}^{\bar{r}\downarrow}&\bar{d}_{L2}^{\bar{y}\downarrow}&\bar{d}_{L2}^{\bar{b}\downarrow} \\
\nu_{\mu,L_1}^\downarrow&c_{R_1}^{r\downarrow}&c_{L_1}^{y\downarrow}&c_{L_1}^{b\downarrow}&\mu^{+\downarrow}_{R1}&\bar{s}_{R1}^{\bar{r}\downarrow}&\bar{s}_{R1}^{\bar{y}\downarrow}&\bar{s}_{R1}^{\bar{b}\downarrow}&\nu_{e,L_1}^\downarrow&u_{R_1}^{r\downarrow}&u_{L_1}^{y\downarrow}&u_{L_1}^{b\downarrow}&e^{+\downarrow}_{R1}&\bar{d}_{R1}^{\bar{r}\downarrow}&\bar{d}_{R1}^{\bar{y}\downarrow}&\bar{d}_{R1}^{\bar{b}\downarrow} \\
\nu_{\mu,L_2}^\downarrow&c_{R_2}^{r\downarrow}&c_{L_2}^{y\downarrow}&c_{L_2}^{b\downarrow}&\mu^{+\downarrow}_{R2}&\bar{s}_{R2}^{\bar{r}\downarrow}&\bar{s}_{R2}^{\bar{y}\downarrow}&\bar{s}_{R2}^{\bar{b}\downarrow}&\nu_{e,L_2}^\downarrow&u_{R_2}^{r\downarrow}&u_{L_2}^{y\downarrow}&u_{L_2}^{b\downarrow}&e^{+\downarrow}_{R2}&\bar{d}_{R2}^{\bar{r}\downarrow}&\bar{d}_{R2}^{\bar{y}\downarrow}&\bar{d}_{R2}^{\bar{b}\downarrow}  \\
\mu^{-\downarrow}_{L1}&s_{L1}^{r\downarrow}&s_{L1}^{y\downarrow}&s_{L1}^{b\downarrow}&\bar{\nu}_{\mu,R1}^\downarrow&\bar{c}_{R1}^{\bar{r}\downarrow}&\bar{c}_{R1}^{\bar{y}\downarrow}&\bar{c}_{R1}^{\bar{b}\downarrow}&e^{-\downarrow}_{L1}&d_{L1}^{r\downarrow}&d_{L1}^{y\downarrow}&d_{L1}^{b\downarrow}&\bar{\nu}_{e,R1}^\downarrow&\bar{u}_{R1}^{\bar{r}\downarrow}&\bar{u}_{R1}^{\bar{y}\downarrow}&\bar{u}_{R1}^{\bar{b}\downarrow} \\
\mu^{-\downarrow}_{L2}&s_{L2}^{r\downarrow}&s_{L2}^{y\downarrow}&s_{L2}^{b\downarrow}&\bar{\nu}_{\mu,R2}^\downarrow&\bar{c}_{R2}^{\bar{r}\downarrow}&\bar{c}_{R2}^{\bar{y}\downarrow}&\bar{c}_{R2}^{\bar{b}\downarrow}&e^{-\downarrow}_{L2}&d_{L2}^{r\downarrow}&d_{L2}^{y\downarrow}&d_{L2}^{b\downarrow}&\bar{\nu}_{e,R2}^\downarrow&\bar{u}_{R2}^{\bar{r}\downarrow}&\bar{u}_{R2}^{\bar{y}\downarrow}&\bar{u}_{R2}^{\bar{b}\downarrow} \\
\mu^{-\downarrow}_{R1}&s_{R1}^{r\downarrow}&s_{R1}^{y\downarrow}&s_{R1}^{b\downarrow}&\bar{\nu}_{\mu,L1}^\downarrow&\bar{c}_{L1}^{\bar{r}\downarrow}&\bar{c}_{L1}^{\bar{y}\downarrow}&\bar{c}_{L1}^{\bar{b}\downarrow}&e^{-\downarrow}_{R1}&d_{R1}^{r\downarrow}&d_{R1}^{y\downarrow}&d_{R1}^{b\downarrow}&\bar{\nu}_{e,L1}^\downarrow&\bar{u}_{L1}^{\bar{r}\downarrow}&\bar{u}_{L1}^{\bar{y}\downarrow}&\bar{u}_{L1}^{\bar{b}\downarrow} \\
\mu^{-\downarrow}_{R2}&s_{R2}^{r\downarrow}&s_{R2}^{y\downarrow}&s_{R2}^{b\downarrow}&\bar{\nu}_{\mu,L2}^\downarrow&\bar{c}_{L2}^{\bar{r}\downarrow}&\bar{c}_{L2}^{\bar{y}\downarrow}&\bar{c}_{L2}^{\bar{b}\downarrow}&e^{-\downarrow}_{R2}&d_{R2}^{r\downarrow}&d_{R2}^{y\downarrow}&d_{R2}^{b\downarrow}&\bar{\nu}_{e,L2}^\downarrow&\bar{u}_{L2}^{\bar{r}\downarrow}&\bar{u}_{L2}^{\bar{y}\downarrow}&\bar{u}_{L2}^{\bar{b}\downarrow} \\
\end{smallmatrix}\right),
\end{equation}
\begin{equation}
\label{eqn:three0}
\mathbb{P}_3\equiv\left(\begin{smallmatrix}
\nu_{\tau,R_1}^\uparrow&t_{R_1}^{r\uparrow}&t_{R_1}^{y\uparrow}&t_{R_1}^{b\uparrow}&\tau^{+\uparrow}_{L1}&\bar{b}_{L1}^{\bar{r}\uparrow}&\bar{b}_{L1}^{\bar{y}\uparrow}&\bar{b}_{L1}^{\bar{b}\uparrow}&\nu_{\mu,R_1}^\uparrow&c_{R_1}^{r\uparrow}&c_{R_1}^{y\uparrow}&c_{R_1}^{b\uparrow}&\mu^{+\uparrow}_{L1}&\bar{s}_{L1}^{\bar{r}\uparrow}&\bar{s}_{L1}^{\bar{y}\uparrow}&\bar{s}_{L1}^{\bar{b}\uparrow} \\
\nu_{\tau,R_2}^\uparrow&t_{R_2}^{r\uparrow}&t_{R_2}^{y\uparrow}&t_{R_2}^{b\uparrow}&\tau^{+\uparrow}_{L2}&\bar{b}_{L2}^{\bar{r}\uparrow}&\bar{b}_{L2}^{\bar{y}\uparrow}&\bar{b}_{L2}^{\bar{b}\uparrow}&\nu_{\mu,R_2}^\uparrow&c_{R_2}^{r\uparrow}&c_{R_2}^{y\uparrow}&c_{R_2}^{b\uparrow}&\mu^{+\uparrow}_{L2}&\bar{s}_{L2}^{\bar{r}\uparrow}&\bar{s}_{L2}^{\bar{y}\uparrow}&\bar{s}_{L2}^{\bar{b}\uparrow} \\
\nu_{\tau,L_1}^\uparrow&t_{R_1}^{r\uparrow}&t_{L_1}^{y\uparrow}&t_{L_1}^{b\uparrow}&\tau^{+\uparrow}_{R1}&\bar{b}_{R1}^{\bar{r}\uparrow}&\bar{b}_{R1}^{\bar{y}\uparrow}&\bar{b}_{R1}^{\bar{b}\uparrow}&\nu_{\mu,L_1}^\uparrow&c_{R_1}^{r\uparrow}&c_{L_1}^{y\uparrow}&c_{L_1}^{b\uparrow}&\mu^{+\uparrow}_{R1}&\bar{s}_{R1}^{\bar{r}\uparrow}&\bar{s}_{R1}^{\bar{y}\uparrow}&\bar{s}_{R1}^{\bar{b}\uparrow} \\
\nu_{\tau,L_2}^\uparrow&t_{R_2}^{r\uparrow}&t_{L_2}^{y\uparrow}&t_{L_2}^{b\uparrow}&\tau^{+\uparrow}_{R2}&\bar{b}_{R2}^{\bar{r}\uparrow}&\bar{b}_{R2}^{\bar{y}\uparrow}&\bar{b}_{R2}^{\bar{b}\uparrow}&\nu_{\mu,L_2}^\uparrow&c_{R_2}^{r\uparrow}&c_{L_2}^{y\uparrow}&c_{L_2}^{b\uparrow}&\mu^{+\uparrow}_{R2}&\bar{s}_{R2}^{\bar{r}\uparrow}&\bar{s}_{R2}^{\bar{y}\uparrow}&\bar{s}_{R2}^{\bar{b}\uparrow}  \\
\tau^{-\uparrow}_{L1}&b_{L1}^{r\uparrow}&b_{L1}^{y\uparrow}&b_{L1}^{b\uparrow}&\bar{\nu}_{\tau,R1}^\uparrow&\bar{t}_{R1}^{\bar{r}\uparrow}&\bar{t}_{R1}^{\bar{y}\uparrow}&\bar{t}_{R1}^{\bar{b}\uparrow}&\mu^{-\uparrow}_{L1}&s_{L1}^{r\uparrow}&s_{L1}^{y\uparrow}&s_{L1}^{b\uparrow}&\bar{\nu}_{\mu,R1}^\uparrow&\bar{c}_{R1}^{\bar{r}\uparrow}&\bar{c}_{R1}^{\bar{y}\uparrow}&\bar{c}_{R1}^{\bar{b}\uparrow} \\
\tau^{-\uparrow}_{L2}&b_{L2}^{r\uparrow}&b_{L2}^{y\uparrow}&b_{L2}^{b\uparrow}&\bar{\nu}_{\tau,R2}^\uparrow&\bar{t}_{R2}^{\bar{r}\uparrow}&\bar{t}_{R2}^{\bar{y}\uparrow}&\bar{t}_{R2}^{\bar{b}\uparrow}&\mu^{-\uparrow}_{L2}&s_{L2}^{r\uparrow}&s_{L2}^{y\uparrow}&s_{L2}^{b\uparrow}&\bar{\nu}_{\mu,R2}^\uparrow&\bar{c}_{R2}^{\bar{r}\uparrow}&\bar{c}_{R2}^{\bar{y}\uparrow}&\bar{c}_{R2}^{\bar{b}\uparrow} \\
\tau^{-\uparrow}_{R1}&b_{R1}^{r\uparrow}&b_{R1}^{y\uparrow}&b_{R1}^{b\uparrow}&\bar{\nu}_{\tau,L1}^\uparrow&\bar{t}_{L1}^{\bar{r}\uparrow}&\bar{t}_{L1}^{\bar{y}\uparrow}&\bar{t}_{L1}^{\bar{b}\uparrow}&\mu^{-\uparrow}_{R1}&s_{R1}^{r\uparrow}&s_{R1}^{y\uparrow}&s_{R1}^{b\uparrow}&\bar{\nu}_{\mu,L1}^\uparrow&\bar{c}_{L1}^{\bar{r}\uparrow}&\bar{c}_{L1}^{\bar{y}\uparrow}&\bar{c}_{L1}^{\bar{b}\uparrow} \\
\tau^{-\uparrow}_{R2}&b_{R2}^{r\uparrow}&b_{R2}^{y\uparrow}&b_{R2}^{b\uparrow}&\bar{\nu}_{\tau,L2}^\uparrow&\bar{t}_{L2}^{\bar{r}\uparrow}&\bar{t}_{L2}^{\bar{y}\uparrow}&\bar{t}_{L2}^{\bar{b}\uparrow}&\mu^{-\uparrow}_{R2}&s_{R2}^{r\uparrow}&s_{R2}^{y\uparrow}&s_{R2}^{b\uparrow}&\bar{\nu}_{\mu,L2}^\uparrow&\bar{c}_{L2}^{\bar{r}\uparrow}&\bar{c}_{L2}^{\bar{y}\uparrow}&\bar{c}_{L2}^{\bar{b}\uparrow} \\
\nu_{\tau,R_1}^\downarrow&t_{R_1}^{r\downarrow}&t_{R_1}^{y\downarrow}&t_{R_1}^{b\downarrow}&\tau^{+\downarrow}_{L1}&\bar{b}_{L1}^{\bar{r}\downarrow}&\bar{b}_{L1}^{\bar{y}\downarrow}&\bar{b}_{L1}^{\bar{b}\downarrow}&\nu_{\mu,R_1}^\downarrow&c_{R_1}^{r\downarrow}&c_{R_1}^{y\downarrow}&c_{R_1}^{b\downarrow}&\mu^{+\downarrow}_{L1}&\bar{s}_{L1}^{\bar{r}\downarrow}&\bar{s}_{L1}^{\bar{y}\downarrow}&\bar{s}_{L1}^{\bar{b}\downarrow} \\
\nu_{\tau,R_2}^\downarrow&t_{R_2}^{r\downarrow}&t_{R_2}^{y\downarrow}&t_{R_2}^{b\downarrow}&\tau^{+\downarrow}_{L2}&\bar{b}_{L2}^{\bar{r}\downarrow}&\bar{b}_{L2}^{\bar{y}\downarrow}&\bar{b}_{L2}^{\bar{b}\downarrow}&\nu_{\mu,R_2}^\downarrow&c_{R_2}^{r\downarrow}&c_{R_2}^{y\downarrow}&c_{R_2}^{b\downarrow}&\mu^{+\downarrow}_{L2}&\bar{s}_{L2}^{\bar{r}\downarrow}&\bar{s}_{L2}^{\bar{y}\downarrow}&\bar{s}_{L2}^{\bar{b}\downarrow} \\
\nu_{\tau,L_1}^\downarrow&t_{R_1}^{r\downarrow}&t_{L_1}^{y\downarrow}&t_{L_1}^{b\downarrow}&\tau^{+\downarrow}_{R1}&\bar{b}_{R1}^{\bar{r}\downarrow}&\bar{b}_{R1}^{\bar{y}\downarrow}&\bar{b}_{R1}^{\bar{b}\downarrow}&\nu_{\mu,L_1}^\downarrow&c_{R_1}^{r\downarrow}&c_{L_1}^{y\downarrow}&c_{L_1}^{b\downarrow}&\mu^{+\downarrow}_{R1}&\bar{s}_{R1}^{\bar{r}\downarrow}&\bar{s}_{R1}^{\bar{y}\downarrow}&\bar{s}_{R1}^{\bar{b}\downarrow} \\
\nu_{\tau,L_2}^\downarrow&t_{R_2}^{r\downarrow}&t_{L_2}^{y\downarrow}&t_{L_2}^{b\downarrow}&\tau^{+\downarrow}_{R2}&\bar{b}_{R2}^{\bar{r}\downarrow}&\bar{b}_{R2}^{\bar{y}\downarrow}&\bar{b}_{R2}^{\bar{b}\downarrow}&\nu_{\mu,L_2}^\downarrow&c_{R_2}^{r\downarrow}&c_{L_2}^{y\downarrow}&c_{L_2}^{b\downarrow}&\mu^{+\downarrow}_{R2}&\bar{s}_{R2}^{\bar{r}\downarrow}&\bar{s}_{R2}^{\bar{y}\downarrow}&\bar{s}_{R2}^{\bar{b}\downarrow}  \\
\tau^{-\downarrow}_{L1}&b_{L1}^{r\downarrow}&b_{L1}^{y\downarrow}&b_{L1}^{b\downarrow}&\bar{\nu}_{\tau,R1}^\downarrow&\bar{t}_{R1}^{\bar{r}\downarrow}&\bar{t}_{R1}^{\bar{y}\downarrow}&\bar{t}_{R1}^{\bar{b}\downarrow}&\mu^{-\downarrow}_{L1}&s_{L1}^{r\downarrow}&s_{L1}^{y\downarrow}&s_{L1}^{b\downarrow}&\bar{\nu}_{\mu,R1}^\downarrow&\bar{c}_{R1}^{\bar{r}\downarrow}&\bar{c}_{R1}^{\bar{y}\downarrow}&\bar{c}_{R1}^{\bar{b}\downarrow} \\
\tau^{-\downarrow}_{L2}&b_{L2}^{r\downarrow}&b_{L2}^{y\downarrow}&b_{L2}^{b\downarrow}&\bar{\nu}_{\tau,R2}^\downarrow&\bar{t}_{R2}^{\bar{r}\downarrow}&\bar{t}_{R2}^{\bar{y}\downarrow}&\bar{t}_{R2}^{\bar{b}\downarrow}&\mu^{-\downarrow}_{L2}&s_{L2}^{r\downarrow}&s_{L2}^{y\downarrow}&s_{L2}^{b\downarrow}&\bar{\nu}_{\mu,R2}^\downarrow&\bar{c}_{R2}^{\bar{r}\downarrow}&\bar{c}_{R2}^{\bar{y}\downarrow}&\bar{c}_{R2}^{\bar{b}\downarrow} \\
\tau^{-\downarrow}_{R1}&b_{R1}^{r\downarrow}&b_{R1}^{y\downarrow}&b_{R1}^{b\downarrow}&\bar{\nu}_{\tau,L1}^\downarrow&\bar{t}_{L1}^{\bar{r}\downarrow}&\bar{t}_{L1}^{\bar{y}\downarrow}&\bar{t}_{L1}^{\bar{b}\downarrow}&\mu^{-\downarrow}_{R1}&s_{R1}^{r\downarrow}&s_{R1}^{y\downarrow}&s_{R1}^{b\downarrow}&\bar{\nu}_{\mu,L1}^\downarrow&\bar{c}_{L1}^{\bar{r}\downarrow}&\bar{c}_{L1}^{\bar{y}\downarrow}&\bar{c}_{L1}^{\bar{b}\downarrow} \\
\tau^{-\downarrow}_{R2}&b_{R2}^{r\downarrow}&b_{R2}^{y\downarrow}&b_{R2}^{b\downarrow}&\bar{\nu}_{\tau,L2}^\downarrow&\bar{t}_{L2}^{\bar{r}\downarrow}&\bar{t}_{L2}^{\bar{y}\downarrow}&\bar{t}_{L2}^{\bar{b}\downarrow}&\mu^{-\downarrow}_{R2}&s_{R2}^{r\downarrow}&s_{R2}^{y\downarrow}&s_{R2}^{b\downarrow}&\bar{\nu}_{\mu,L2}^\downarrow&\bar{c}_{L2}^{\bar{r}\downarrow}&\bar{c}_{L2}^{\bar{y}\downarrow}&\bar{c}_{L2}^{\bar{b}\downarrow} \\
\end{smallmatrix}\right).
\end{equation}
%
%


\bibliography{Jordanalgebra}  
\bibliographystyle{unsrt}

\end{document}